\documentclass[12pt]{article}
\newcommand{\bm}[1]{\mbox{\boldmath$#1$}}
\newcommand{\X}{{\mathbf X}}
\newcommand{\Z}{{\mathbf Z}}
\newcommand{\0}{{\mathbf 0}}
\usepackage[flushleft]{threeparttable}

\newcommand{\bbeta}{\mbox{\boldmath$\beta$}}
\newcommand{\bgamma}{\mbox{\boldmath$\gamma$}}
\newcommand{\balpha}{\mbox{\boldmath$\alpha$}}
\newcommand{\bmu}{\mbox{\boldmath$\mu$}}
\newcommand{\btau}{\mbox{\boldmath$\tau$}}

 \usepackage{xr}
 \usepackage{zref}
\usepackage{adjustbox}

\RequirePackage[OT1]{fontenc}
\usepackage{authblk}
\RequirePackage{amsthm,amsmath}
\RequirePackage[]{natbib}
\RequirePackage[colorlinks,citecolor=blue,urlcolor=blue]{hyperref}
\usepackage{rotating}
\usepackage{array}
\usepackage[]{natbib}
\usepackage{caption}
\usepackage{latexsym}
\usepackage{amsmath}
\usepackage{multirow}
\usepackage{graphicx}
\usepackage{graphics}
\usepackage{amssymb}
\usepackage{epstopdf}
\usepackage{booktabs}
\usepackage{subfig}
\usepackage{placeins}
\RequirePackage{xr,zref}

\usepackage{setspace}
\author[1]{Andriy Derkach}
\author[1,2]{Ruth Pfeiffer}
\affil[1]{Division of Cancer Epidemiology and Genetics, National Cancer Institute, National Institutes of Health, Rockville MD 20850, USA}
\affil[2]{Corresponding author: pfeiffer@mail.nih.gov}

\begin{document}
	
		\title{Subset Testing and Analysis of Multiple Phenotypes (STAMP)}
		\date{}	
	\maketitle	
		\begin{abstract}
				Meta-analysis of multiple genome-wide association studies (GWAS) is  effective  for detecting single or multi marker associations with complex traits.
				We develop a flexible procedure (``STAMP'') based on mixture models to  perform    region based  meta-analysis of 
				different  phenotypes using data from different GWAS and identify subsets of associated phenotypes. 
				Our model framework helps  distinguish  true associations from between-study heterogeneity.
				As a measure of association we  compute for each phenotype the  posterior probability that  the genetic region under investigation is truly associated.  Extensive simulations show that STAMP is more powerful than standard approaches for meta analyses when the proportion of truly associated outcomes is $\leq$ 50\%.   For  other settings, the power of STAMP is similar to that of existing methods. We illustrate our method on two examples,  the association of a region on chromosome 9p21 with risk of fourteen cancers, and the associations of expression of quantitative traits loci (eQTLs) from two genetic regions with their cis-SNPs  measured in  seventeen  tissue types using data from The Cancer Genome Atlas (TCGA).
		\end{abstract}

	\section{Introduction}
	\label{sec1} Sometimes it is of interest to assess the association of genetic variation within a pre-specified region  with different, possibly related, phenotypes, and to quantify heterogeneity of the associations.  
	For example, \citet{Li2014} recently studied the associations of  single nucleotide polymorphisms (SNPs) in
	a chromosome 9p21 region with eight cancers  that includes interferon genes and several tumor suppressor genes, from eight genome-wide association (GWAS) studies.
	The authors conducted SNP-level analyses for each cancer and used a subset-based statistical approach (ASSET) \citep{Bhattacharjee2012} to combine SNP-level p-values across   cancers. 
	In another example,  \citet{flutre2013statistical} proposed methods to assess single SNP associations between   with  expression quantitative trait loci  (eQTL) expression measured in multiple tissues.
	
	Standard meta-analytic approaches to combine summary information from a single SNP are not powerful when the SNP has an effect in only a subset of phenotypes or in opposite directions for some phenotypes.  Multiple methods are available to assess the association of common genetic variants such as GWAS SNPs with risk of multiple phenotypes measured on the same samples \citep{fisher,Multi,Sluis2013} but only few methods are available based on summary statistics. ASSET and CPBayes \citep{Majumdar2017} use summary statistics to identify subsets of studies associated with a particular SNP, but they do not allow one to readily combine information from multiple SNPs in a locus. Information stemming from linkage disequilibrium (LD) is not utilized when analyzing each SNP in a region separately. Several adaptive gene-based approaches   are available to study multiple SNPs   simultaneously \citep{Tang2012,Sluis2015,Kwak2017} and  accommodate heterogeneous SNP effects,  or effects that present in some studies. However,  these approaches only give global measures of association and do not identify the subset of  associated studies.
	
	We therefore propose a new  approach  to explore  genetic heterogeneity of   associations for a genomic region with different phenotypes and to  identify a subset of phenotypes that are   associated  with that  region. First, for each phenotype separately, we combine  the SNP specific association estimates  using an aggregated level test statistic. We then  assume that the test statistics  arise  from a mixture  distribution with two components, one under the null model of no association of the study specific phenotype with the  genetic region,  and one distribution assuming that there is an association.  We use  a hierarchical model to describe SNP   effects (Section 2) that can accommodate varying levels of between-phenotype  heterogeneity.
	We then  test if the mixture distribution provides a better fit to the region specific test statistics from all studies than a single component density, estimate the parameters of the mixture and compute posterior probabilities that  a particular phenotype is   associated with the genomic  region (Section 3).  As an illustration, we  analyzed the association of  the 9p21 region (using GWAS SNPs) with various cancers,  and 
	the genetic  associations of eQTLs from two genetic regions measured in  seventeen different tissue types  (Section 4).   We study our method in simulations (Section 5) and compare its power to existing meta analytic approach, before closing with a discussion (Section 6).

	\section{Data and models}
	\subsection{Association models}
	
	We now describe the model assumed to govern the association between a particular phenotype $Y_s$ and  
	genotypes $\X_s=(X_{s1},\ldots,X_{sp_s})'$ for $p_s$ SNPs   measured  in a genomic region,  where $X_{si} =0,1,$ or $2$ denotes  the number of minor alleles at locus $i,$ $i=1, \ldots, p_s$. Here we allow for different numbers of SNPs measured in a genomic region for different phenotypes. We consider the generalized linear model (GLM) setting   \citep[see e.g.][]{mccullagh89},
	and assume that the conditional expectation   of  $Y_s$ given $\X_s$ is
	\begin{equation}
	E_F(Y_s |   \X_s,\bgamma_s) = h(\gamma_{s0}+\sum_{i=1}^{p_s} \gamma_{si} X_{si}) = h(\gamma_{s0}+\bgamma_s' \X_s), 
	\label{model}
	\end{equation}
	where $h$ is  a known   function and $(\gamma_{s0}, \bgamma_s)$ 
	a vector of $p+1$ association parameters for the SNP. If the k-th SNP is not  associated with  $Y_s$, $\gamma_{sk}=0$. Additional  covariates $\Z_s=(Z_{s1},\ldots,Z_{sq})'$ can easily be accommodated in model (\ref{model})
	through  $ E_F(Y_s |  \X_s,\gamma_{s0},\bgamma_s,\balpha_s) = h(\gamma_{s0}+\bgamma_s' \X_s + \balpha_s' \Z).$  
	We assume that $F$ is  a probability density or mass function  from the exponential family  \citep{mccullagh89}. 

	\subsection{Properties of estimates obtained from   marginal SNP models}
	
	In GWAS studies
	the  estimate  $\hat \beta_{sj}$ for the association of the jth  SNP  with outcome $Y_s$ is typically  obtained by maximizing  a marginal likelihood    that only includes the genotype $X_{sj}$ for the jth SNP in the specification of the mean function instead of the whole vector $\X_s$, 
	\begin{eqnarray}
	\label{M4}
	E_G(Y_s  |   \X_s,\beta_{s0j}, \beta_{sj}) =  h(\beta_{s0j}  + \beta_{sj}X_{sj}),
	\end{eqnarray}
	where $h$ denotes the same   function as in (\ref{model}). If additional covariates $\bm Z_s$ are available, (\ref{M4}) can be extended to $E_G(Y_s  |   \X_s,\beta_{s0j}, \beta_{sj},\bm \zeta_s) =  h(\beta_{s0j}  + \beta_{sj}X_{sj} +\bm \zeta'_{sj} \bm Z_s)$. We use the subscript $G$ to denote the  misspecified marginal mean  probability model   that uses only individual   SNP genotypes.
	We show in the Appendix  \ref{sec:bias}  that, conditional on $\bgamma_s$,  the estimate $\hat{\beta}_{sj}$ based on (\ref{M4}) converges to $\beta_{sj}$ that satisfies the equation 
	\begin{equation}
	\beta_{sj}  =   \frac{\sum_{i=1}^{p_s} \gamma_{si}  Cov(X_{si},X_{sj}) }{  Var(X_{sj}) }, 
	\label{eq: marginalvsfull}
	\end{equation} 
	where $\gamma_{si}$  is the true associate parameter for SNP $i$ in (\ref{model}), when $h$ is   the identity link function $h$ or the logistic link,  under both,  prospective and retrospective sampling, i.e. for case-control data assuming rare disease.
	As can be seen directly from (\ref{eq: marginalvsfull}), when there is no association, i.e. $\gamma_{si}=0$ for all SNPs $i=1,...,p_s$, then also $\beta_{si}=0$ for all $i=1,...,p_s$, and when the SNPs are uncorrelated, then $\beta_{si}=\gamma_{si}$. Using the matrix $\Omega_s$ defined element-wise as    
	
	 \begin{equation} 
	 \label{omega}
	[\Omega_s]_{ij}  = \frac{Cov(X_{si},X_{sj}) }{  Var(X_{si}) } \text{ for } i,j = 1,...,p_s,
	\end{equation}
	and conditional on the vector of true effects $\bm \gamma_s$, the estimates $\hat{\bm \beta_s}=(\hat \beta_{s1},\ldots,\hat \beta_{sp_s})$ from the marginal model (\ref{M4}) have the following limiting distribution  
	\begin{equation}
	\hat{\bm \beta_s}|\bm \gamma_s \sim N(\bm \beta_s, \Sigma_s) = N(\Omega_s\bm \gamma_s, \Sigma_s),
	\label{eq: 9}
	\end{equation}
	where $\Sigma_s = Cov(\hat{\bm \beta_s}|\bm \gamma_s)$, which is typically not known for the marginal estimates. For small effects $\bm \gamma_s$, following \cite{Hu2013},

	\begin{equation}
	\Sigma_s \approx D_s \Upsilon_s D_s,
	\label{eq: vars}
	\end{equation}
	where $\Upsilon_s=Cor(X_s)$ is the correlation matrix between the $p$ SNPs that is assumed to be known and  $D_s=diag(\sigma_{s1},\ldots,\sigma_{sp_s})$ is a diagonal matrix of standard error estimates  $\sigma_{sj}$  of   $\hat{\beta}_{sj}$, $j=1,...,p_s$. 
	Letting $V_{\X_s}=diag\left\{Var(X_{s1}),\ldots,Var(X_{sp_s})\right\}$, 
	$\Upsilon_s=V_{\X_s}^{1/2} \Omega_s V_{\X_s}^{-1/2} $. \citet{Visscher2012} derived similar results to (\ref{eq: marginalvsfull}) using a least squares approach for the linear model and extended it to case-control data using a liability threshold model.

	\subsection{Hierarchical model for SNP effects} 
	
\begin{figure}[htp!]
		\begin{center}
			$
			\includegraphics[scale=0.85]{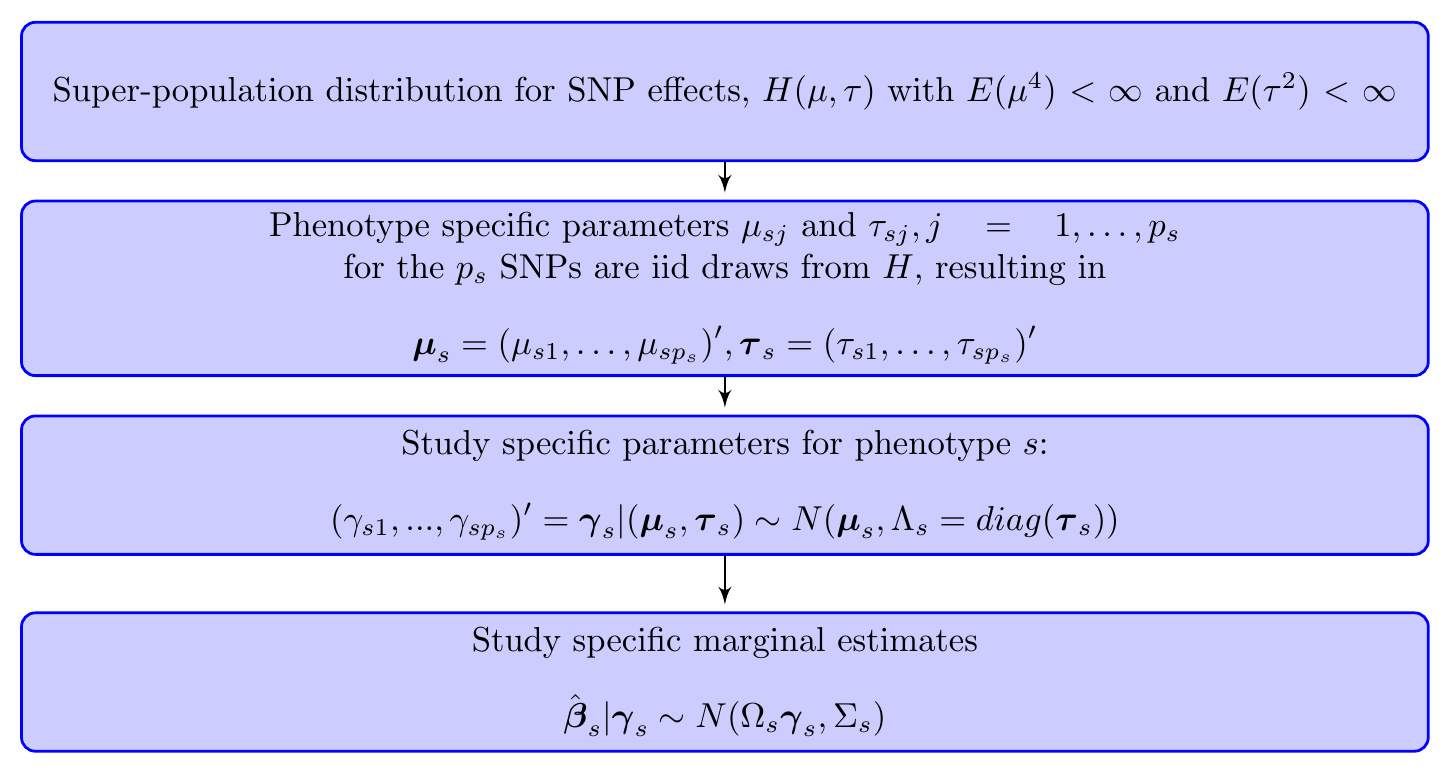}
			$
		\end{center}
		\centering
		\caption{Hierarchical model set-up for  study specific estimates for particular phenotype $s$}
		\label{Figure1}
	\end{figure}

	We assume that   the study and phenotype specific  association parameters $ \bm \gamma_{s}$ in (\ref{model})   arise from a multivariate normal distribution,
	\begin{equation}    
	\bm \gamma_{s} = (\gamma_{s1},...,\gamma_{sp_s})' \sim N(\bm \mu_{s},\Lambda_s),  s=1,\ldots, S,
	\label{eq: her}
	\end{equation}
	where 	$\bm \mu_{s} = (\mu_{s1},\ldots,\mu_{sp_s})' $ and $\bm \tau_s=(\tau_{s1},\ldots,\tau_{sp_s})'$  denote the phenotype specific  association parameters  and $\Lambda_s = diag(\bm \tau_s)$ is a diagonal matrix. 
The components $\mu_{sj}$ and $\tau_{sj}$,  $j=1,\ldots,p_s,$ of  $\bm \mu_{s}$ and $\bm \tau_s$,
  are assumed to be independent random draws from two possible super-populations, one for associated phenotypes and one for phenotypes that exhibit no associations with the region (Figure \ref{Figure1}). 
 We do not assume any specific distributions for the super-populations, we only describe them through their moments. 





We distinguish between phenotype specific mean SNP effects $\bmu_s$  and study specific effects $\bgamma_s$ 
as different studies for the same genotype could have different ''true'' associations, e.g. due to differences in unmeasured confounders.  
If there are multiple studies for each phenotype, then $(\bmu_s, \btau_s)$ can be estimated from available data. Otherwise, only the super-population parameters in the top hierarchical layer can be estimated.

Based on equations  (\ref{eq: 9}) and (\ref{eq: her}),  the conditional distribution of $\hat{\bm \beta_s}$   is 
\begin{equation}
\hat{\bm \beta}_s|(\bm \mu_s, \bm \tau_s) \sim N\left(\Omega_s\bm \mu_s,  \Omega_s\Lambda_s\Omega_s' + \Sigma_s\right),
\end{equation}  
where $\Sigma_s$ is given in (\ref{eq: vars}).
To recover the true association parameters $\bm \mu_s$ and $\Lambda_s$, we rotate the estimates, to obtain
\begin{equation}
\hat{\bbeta}^*_s|(\bm \mu_s, \bm \tau_s)  = \Omega_s^{-1}\hat{\bm \beta}_s |(\bm \mu_s, \bm \tau_s) \sim N\left( \bm \mu_s,  \Lambda_s +  \Sigma_s^*\right), 
\label{distbeta}
\end{equation}  
where $ \Sigma_s^*= \Omega_s^{-1}\Sigma_s\Omega_s'^{-1}$.
Under local alternatives, i.e. small effects $\bm \gamma_s$,  \citet{Tang2014,Visscher2012}  showed that  $\Sigma^*_s \approx  \frac{c_s}{N_s} Cov(\bm X)^{-1}$, where $c_s$ denotes the residual variance under the null model of no genetic associations,  and $N_s$ is the sample size of study $s$.

For those  $Y_s$ that exhibit no associations with the genetic region, we consider two different assumptions for the super-population that gave rise to  $\bmu_s$ and $\btau_s$, termed ``null models''. 
Under the first one,  the ``strong null model ($m_0^{st}$)'',   that has been used  for meta analysis of single or multiple SNPs \citep[]{Eskin2011,Lee2013,Tang2014,Shi2016}, $\bm \mu_s \equiv \bm 0$,    and $\bm \tau_s \equiv \bm 0$  for all SNPs in a region, and thus   $\gamma_{sj} \equiv 0$,  $j=1,...,p_s$,  without any variation. Thus the first three levels in the hierarchical model in Figure \ref{Figure1} can be collapsed, and it follows that  
$\hat{\bbeta}_s|(\bm \mu_s, \bm \tau_s) \stackrel{m_0^{st}}{ \sim}  N\left( \0, \Sigma_s\right). 
$
Several  super-population models are appropriate when $m_0^{w}$ is used for those $Y_s$ for which there are associated SNPs in the  genetic region. The first   is to assume that $E(\mu_{sj})= 0$ and $Var(\mu_{sj})+Var(\tau_{sj})\neq 0$. This setup has been used previously for variance component testing in random effect models \citep[e.g.][]{LIN1997} and for het-SKAT\citep[]{Lee2013}.
 Alternatively, one could let $E(\mu_{sj})\neq 0$ and $Var(\mu_{sj})+Var(\tau_{sj})= 0$, which is assumed in fixed effect meta analysis \citep[]{cochran1954combination}.
\cite{Eskin2011,Tang2014} studied a  combination of two models,  $E(\mu_{sj})\neq0$ or $Var(\mu_{sj})+E(\tau_{sj})\neq 0$. 

Under the second, weaker  set of assumptions for the null super-population model $(m_0^w)$, we only assume that   $ \mu_j \equiv 0$  for all SNPs $j$.  Thus, under $m_0^{w}$, for some SNPs,  $\gamma_{sj}\neq 0$   due to different LD in different populations, measurement error or other sources of confounding.  
The appropriate model for phenotypes with associations in the region, that has been used in the context of meta-analysis \citep[e.g.][]{Eskin2011,Tang2014,Shi2016},  assumes that $E(\mu_{sj}) \neq 0$ or  $Var(\mu_{sj}) \neq 0$. Here, we require the availability of a  'negative' control study, i.e. a phenotype $Y_s$ that is known not to be associated  with the genetic region, to be able to distinguish between sources of variation in  the genetic effects, i.e. between $Var(\mu_{sj})$ and  $E(\tau_{sj})$.

To summarize, the distributions of the rotated estimates of effect sizes in (\ref{distbeta})   simplify to
\begin{equation}\hat{\bbeta}^*_s|(\bm \mu_s, \bm \tau_s)  \stackrel{m_0^{st}}{ \sim}  N\left( \0, \Sigma_s^*\right)
\mbox{ and } \hat{\bbeta}^*_s|(\bm \mu_s, \bm \tau_s)   \stackrel{m_0^{w}}{ \sim}  N\left( \0, \Lambda_s + \Sigma_s^*\right),
\label{distbetanull}
\end{equation}  
under the two models of no genetic associations.

\section{Assessing the association of  a genetic region with  multiple phenotypes}

We assume now that we have one study for each phenotype $Y_s$. 
For each phenotype  $Y_s$ we combine the linearly transformed values $\Sigma^{*-1}_s\hat{\bbeta^*_s}$  using   a linear or quadratic statistic
$T_s$, which are asymptotically  equivalent to variance component tests to assess high dimensional alternatives \citep[]{derkach2014,Tang2014,Lee2012}

Linear tests have good power if a large proportion of SNPs in the  region under consideration are associated and have effects in the same direction, while quadratic test statistics are robust to different signs of effect estimates and are more powerful when the proportion of associated SNPs in the  region is small \citep[e.g.][]{derkach2014}.
Under heterogeneity of associations of phenotypes $Y_s, s=1,\ldots, S$, we assume
that $T_s$ arises from  a mixture model that we present next.

\subsection{Mixture model}
	\label{Sec 3.1}
If only a proportion of the phenotypes $Y_s, s=1,\ldots,S,$ are associated with the genetic region under investigation, we assume test statistics $T_s$   arise from a mixture  distribution, due to two super populations giving rise to the observed estimates,
\begin{equation}
f(T_s) \sim (1-\pi) \varphi_0 (T_s)+\pi \varphi_1(T_s).
\label{mixture}\end{equation}

In (\ref{mixture}), $\varphi_0$ denotes the density  of $T_s$  under the null model  of no association of that particular genetic region with $Y_s$, and $\varphi_1$ is the density when the region is  associated with the phenotype. The mixing proportion $\pi$ can be interpreted as the prior probability of a phenotype  having genetic associations. Functional information can be incorporated into $\pi$, e.g.  by using a covariate $Z_s$ that captures biologically relevant data through $\pi_s = \exp(\delta_0+\delta_1 Z_s)/\left\{1+ \exp( \delta_0+\delta_1  Z_s)\right\}$. 

For both, our linear and quadratic summary statistics $T_s$, $\varphi_0$ and  $\varphi_1$   can be  approximated by  normal densities. 
We  discuss the parameterization of $E_i(T_s)$  and $Var_i(T_s), i=0,1,$ and the estimation of model (\ref{mixture})   in detail in  Sections \ref{sec:TL} and \ref{sec:TQ}, respectively and summarize it in Table \ref{table 1b}.

\begin{table}[htbp]
	\centering
	\caption{{Summary of models and parameters estimated under a single component model, $\varphi_0$ ($H_0$,), or a mixture distribution ($H_1$) that indicates heterogeneity of associations. We let $ \mu_c = \mu - E(\mu)$. $T^L$ and $T^Q$ denote linear and quadratic test statistics under model (\ref{mixture})}}
	\begin{tabular}{p{1.5cm} p{3.0cm}p{7.5cm}}
		&\multicolumn{2}{c}{{\bf Parameters estimated under}}\\
		{\bf Test Statistic} &  $H_0$ (\textbf{single density}) & $H_1$ ({\bf mixture, i.e. heterogeneity}) \\
		\hline
		\\
		\multicolumn{3}{c}{"Weak" null model: $\bm \mu\equiv \0$,   $ \hat{\bbeta}^*_s|\bm \mu_s, \bm \tau_s  \sim  N\left( \0,\Lambda_s+  \Sigma_s^*\right)$ } \\[10pt]
		\hline
		\\[2pt]
		$T^L_{}$ & $E(\tau)$ & $\pi, E(\mu),E(\tau),Var(\mu)$   \\[8pt]
		$T^Q_{}$ & $E(\tau),Var(\tau)$ & $\pi, E(\mu),E(\mu^2_c),E(\mu^3_c),E(\mu^4_c),E(\tau),Var(\tau)$  \\[8pt]
		\hline
		\\
		\multicolumn{3}{c}{"Strong" null model: $\bm \mu\equiv \0$ and $\bm \tau\equiv \0$,   $ \hat{\bbeta}^*_s|\bm \mu_s, \bm \tau_s  \sim  N\left( \0,\Sigma_s^*\right)$ } \\[10pt]	
		\hline
		\\[2pt]
		$T^L_{}$ & $-$ & $\pi,E(\mu),e^\zeta=E(\tau)+Var(\mu)$  \\[8pt]
		$T^Q_{}$ & $-$ & $\pi,e^\zeta = E(\tau) + E(\mu^2_c)$, $E(\mu),E(\mu^3_c)$,
		
		$e^\psi =3Var(\tau) + Var(\mu^2_c) + 2E(\tau)^2 + 4E(\tau) E({\mu^2_c})$ \\[8pt]
		\hline
	\end{tabular}%
	\label{table 1b}%
\end{table}

The basic steps for assessing heterogeneity of associations for phenotypes $Y_s,~s=1,\ldots,S,$  and for identifying  the subset of phenotypes associated with a genomic region are as follows.  
\begin{enumerate}
	\item
Use a likelihood ratio test (LRT) to test if the statistics $T_s, s=1,\ldots,S$, arise from the mixture model in (\ref{mixture}) ($H_1$), or from a single density, $\varphi_0$ ($H_0$). For testing under the weak null model, a "control" study, i.e. a study under $m_0^w$ is required for identifiability. Details on the testing are given in Section \ref{sex:test}. 
	
	\item If there is evidence of heterogeneity based on the LRT, use the mixture model to compute the probability that the region is associated with  a  particular phenotype  $Y_s$, i.e.  the posterior probability 
	$$\hat p_s= P(T_s \mbox{ arises from } \varphi_1 |T_1,\ldots,T_S) = \frac{\hat \pi \hat\varphi_1 (T_s)}{(1-\hat \pi) \hat \varphi_0 (T_s)+\hat \pi \hat \varphi_1 (T_s)}.$$
	\item If 
	$\hat p_s>p^*$ for some prespecified threshold, e.g. $p^*=0.5$,  then phenotype $Y_s$ is considered to be associated with the region. 
\end{enumerate}

\subsection{A linear summary test  statistic, $T_s^L$}
\label{sec:TL}

We first  propose and study a  linear test statistic to combine transformed SNP effects, 
\begin{equation}
T^L_s = \frac{\bm 1'\Sigma^{*-1}_s \hat{\bbeta}^*_{s}}{\left\{\bm 1'diag(\Sigma^{*-1}_s)\bm 1\right\}^{1/2}}.
\label{linear}
\end{equation}
As $\hat{\beta}^*_{sj}$ is asymptotically normally distributed,  $T^L_s$ conditional on $\bm \mu_s$ and $\Lambda_s$ is normally   distributed with moments
\begin{equation}
E_{\bm \mu_s,\bm \tau_s}(T^L_s) =\frac{\bm 1'\Sigma^{*-1}_s\bm \mu_s}{\left\{\bm 1'diag(\Sigma^{*-1}_s)\bm 1\right\}^{1/2}} \mbox{ and } 
Var_{\bm \mu_s,\bm \tau_s}(T^L_s) = \frac{\bm 1' \Sigma^{*-1}_s \bm 1+\bm 1' \Sigma^{*-1}_s\Lambda_s\Sigma^{*-1}_s \bm 1}{\left\{\bm 1'diag(\Sigma^{*-1}_s)\bm 1\right\}^{1/2}}. 
\end{equation}
The unconditional mean and variance of $T^L_s$ are
\begin{equation}
E(T^L_s) =\frac{\bm 1'\Sigma^{*-1}_sE(\mu)}{\left\{\bm 1'diag(\Sigma^{*-1}_s)\bm 1\right\}^{1/2}} \mbox{ and } 
Var(T^L_s) = \frac{\bm 1' \Sigma^{*-1}_s \bm 1+\left\{E(\tau)+Var(\mu)\right\}\bm 1' \Sigma^{*-2}_s \bm 1}{\left\{\bm 1'diag(\Sigma^{*-1}_s)\bm 1\right\}^{1/2}}. 
\label{linearmoments}
\end{equation}
The numerator of the variance of  $T^L$ under the strong null  model is  
$\bm 1' \Sigma^{*-1}_s \bm 1$ 
and under the weak null model  it is 
$\bm 1' \Sigma^{-1*}_s \bm 1+E(\tau)\bm 1' \Sigma^{*-2}_s \bm 1$.
Under the alternative model  	$E(T^L_s)$ and $Var(T^L_s)$ in (\ref{linearmoments}) do not simplify further. 
For the LRT based on  the weak null model we estimate four parameters under the alternative model and one under the null model (see Table 1).  


\subsection{A quadratic summary test statistic, $T_s^Q$}
\label{sec:TQ}

The linear test statistic $T^L_s$ in   Section 3.2   has the disadvantage that it is sensitive to the directions of the associations, i.e. the signs of the $\beta^*_{si}, i=1,\ldots, p_s$, and is not powerful when signal comes from only a few SNPs. 
To overcome these limitations we also  combine the $p_s$ SNP  estimates for  phenotype  $s$   using a quadratic form,
\begin{equation}
T^Q_s = \hat{\bbeta}^{*'}_{s}\Sigma^{*-1}_sW_s\Sigma^{*-1}_s\hat{\bbeta}^*_{s},  
\label{eq: teststat}
\end{equation} 
where $W_s$ is a preselected weight matrix. Since the $\hat{\bbeta}^*_{s}$ have an asymptotically multivariate normal distribution, $T^Q_s$ is a linear combination of independent non-central chi-squared random variables \citep{derkach2014,wu} where the non-centrality parameters depend on   $\bm \mu_s$ and $\bm \tau_s$.
Within the normal mixture framework in Section \ref{Sec 3.1} we utilize that if the number $p_s$  of SNPs is large,  $T^Q_s$ is  approximately  normally distributed  with mean $E(T^Q_s)/\sqrt{p_s}$  and variance  $Var(T^Q_s)/{p_s}$. Note that for $W_s=\Sigma^*_s$, $T^Q_s$ corresponds to the   Hotelling's test  statistic \citep{derkach2014,Tang2014}. Here, we let $W_s=I$, where $I$ denotes identity matrix. This choice  may improve power because it assigns bigger weights to the largest principal components of  $Cov(\bm X)$, which are likely to explain a large proportion of the phenotypic variation. For small $\gamma_i$, $T_s^Q$  is asymptotically equivalent to the C-alpha test  for rare variants under local alternatives \citep{Neale2011}. Other choices of $W_s$ based on 
MAFs were proposed in \citet{wu} and \citet{Basu2011} in the context of rare variant analysis.

Based on the conditional moments given  in Appendix \ref{sec:moments},  the unconditional moments are 
\begin{equation}
E(T^Q_s) = 
tr(\Sigma^{*-1}_s) + e^\psi tr(\Sigma_s^{*-2}) + \left\{E(\mu)\right\}^2 \bm 1'\Sigma_s^{*-2}\bm 1,
\label{eq24}
\end{equation}
where  $e^\psi = Var(\mu) + E(\tau)$ quantifies the variability in genetics effects due to within locus and between study heterogeneity and $e^\zeta = 3Var(\tau)+Var\left[\left\{\mu - E(\mu)\right\}^2\right] + 2\left\{E(\tau)\right\}^2 + 4Var(\mu)E(\tau)$ is used to capture the higher order moments of the super-population. Letting $\mu_c=\mu-E(\mu)$,
\begin{multline}
Var(T^Q_s) =2tr(\Sigma_s^{*-2}) + 4tr(\Sigma_s^{*-3})e^\psi  + 4\bm 1' \Sigma_s^{*-4} \bm 1\left\{E(\mu)\right\}^2e^\psi  \\ + 2tr(\Sigma_s^{*-4})e^{2\psi} + 4\left\{E(\mu_c)\right\}^3E(\mu)\bm 1'\Sigma_s^{*-2}diag(\Sigma_s^{*-2}) \\  + tr\left\{\Sigma_s^{-2}diag(\Sigma_s^{*-2})\right\}\left(e^\zeta -2e^{2\psi}\right) + 4\bm 1' \Sigma_s^{*-3} \bm 1\left\{E(\mu)\right\}^2.
\label{eq25}
\end{multline}
In summary, (\ref{eq24}) and  (\ref{eq25}) depend on the following moments of the distribution of $\bm \mu_s$ and $\bm \tau_s$:  $E(\mu)$, $E(\mu^3_c)$, $E(\mu^2_c)=Var(\mu)$, $E(\mu^4_c)$, $E(\tau)$ and $Var(\tau)$ (see Table \ref{table 1b}). The moments of  $T^Q_s$ for a general matrix $W_s$ are given in the Appendix \ref{sec:moments}. Under the strong null model, (\ref{eq24}) and (\ref{eq25}) simplify to $E_0(T^Q_s) = tr(\Sigma_s^{*-1}) \text{ and } Var_0(T_s^Q) = 2 tr(\Sigma^{*-2})$, and under the weak null model to
$E_0(T^Q_s) = tr(\Sigma_s^{*-1}) + E(\tau)tr(\Sigma_s^{*-2})
$
~and 
$Var_0(T^Q_s) =2tr(\Sigma_s^{*-2}) + 4tr(\Sigma_s^{*-3})E({\tau})$ $+ 2tr(\Sigma_s^{*-4})E({\tau})^2 +3tr\left\{\Sigma_s^{-2}diag(\Sigma_s^{*-2})\right\}Var(\tau).
$




The identifiability of the parameters   in the first two moments of $T^Q_s$ under either  null model can be seen immediately. Here, we thus discuss identifiability of $\zeta$, $\psi$, $E(\mu)$ and $E(\mu^3_c)$ from (\ref{eq24}) and (\ref{eq25}) under the model for association. 
The signs of $E(\mu)$ and $E(\mu^3)$ are not identifiable. 
The identifiability of $\psi$ and $E(\mu)^2$ is ensured  from  the form of $E(T_s^Q)$ if there are at least two studies with different  matrices $\Sigma^{*-2}_s$. Similarly $E(\mu_c^3)$ and $\zeta$ are identifiable from the second moments of $T^Q_s$ if there are at least two studies with different  matrices $\Sigma^{*-2}_sdiag(\Sigma^{*-2}_s)$. If $tr(\Sigma^{*-2}_s) =  \bm 1'\Sigma_s^{*-2}\bm 1$ (e.g. the SNPs are independent),  we cannot distinguish between effects of $E(\mu^2)$ and $E(\tau)$. This special case is further discussed in Appendix \ref{sec:indep_case}.


\subsection{Testing for heterogeneity of associations among studies}
\label{sex:test}

Testing for heterogeneity of associations among phenotypes $Y_s$ with the proposed statistics corresponds to assessing if   $T_s^L$ or $T_s^Q$   arise from a single density or a mixture of densities. We thus use a LRT statistic for $T_{s}^L$ or $T_{s}^Q$ and propose two parametric bootstrap procedures to compute p-values, one  for the strong and one under the weak null model. 

For testing under  the strong null model,  for each bootstrap replication $r$, we generate rotated estimates $\hat{\bm \beta}^*_s(r) \sim MVN(\bm 0,\Sigma^*_s)s$ for  $s=1,...,S$. Then we recalculate the test statistic and obtain a new value of  $LRT(r)$ based on the vector of  $T^Q_s(r)$   or $T^L_s(r)$. When testing with  the  weak null model, however, the replication procedure is more complicated, as the distribution of the marginal estimates depends on the   diagonal matrix $\Lambda_s$, i.e. the second moment of the $\tau_i$.  We consider two different procedures for $T^Q_{s}$ and $T^L_{s}$.   For the linear statistic, we directly generate $T^L_s(r)$ from a normal distribution with mean $0$ and covariance matrix $ \left\{\bm 1'\Sigma^{*-1}_s \bm 1 + \hat E(\tau) \bm 1' \Sigma^{*-2}_s \bm 1\right\} /\left\{\bm 1'diag(\Sigma_s^{*-1})\bm 1\right\}^{1/2}$, where $\hat E(\tau) $ is estimated from moments of the linear statistic (\ref{linearmoments}).  

We do not generate $T^Q_s$ directly from a normal distribution, because when $p_s$, the number of SNPs is small,  or LD is high in  the region, the normal approximation may not be appropriate. Instead, we generate the estimates of the effect sizes as functions of $\tau$ as follows. We estimate $E(\tau)$ and $E(\tau^2)$ by solving  two unbiased estimation equations under the restriction that the estimates cannot be negative, 
\begin{equation}
\sum_{s=1}^S\sum_{j=1}^{p_s} \left\{\frac{\hat{\beta}^2_{sj}}{\sigma^2_{sj}} - 1 - \frac{E({\tau})}{\sigma^2_{sj}} \right\} =0
\text{ and }
\label{eq: m1}
\sum_{s=1}^S\sum_{j=1}^{p_s} \left\{\frac{\hat{\beta^4}_{sj}}{\sigma^4_{sj}} - 1- 3\frac{E({\tau^2})}{\sigma^4_{sj}} - 6\frac{E({\tau})}{\sigma^2_{sj}} \right\}=0, 
\end{equation}
to obtain  $\hat V(\tau) = \max\left\{0,\hat E(\tau^2) - \hat E(\tau)^2\right\}$.  
We then 
draw the elements of the diagonal matrix $\Lambda_s(r)$   from an inverse-gamma distribution with the first two moments equal to $\hat{E}(\tau)$ and $\hat{E}(\tau^2)$, generate transformed marginal estimates 
$\hat{\bm \beta}^*_s(r) \sim N\left\{\bm 0,\Sigma^*_s + \Lambda_s(r)\right\}$ and calculate the quadratic statistics $T_s^Q(r)$.

For both procedures, p-values are calculated as 
$	\hat{p} = 1/R \sum_{r=1}^{R} I\left\{LRT(obs)\leq LRT(r)  \right\}$, where $I$ denotes the indicator function.

\section{Data examples}

We  illustrate our method on two data examples, one that uses binary phenotypes and one based on  continuous $Y_s$.

\subsection{Association of a chromosome 9p21 region with  multiple cancers}

We used  data from GWAS studies in dbGaP to  assess the association of a region on chromosome 9p21  with  fourteen different cancers (see Supplemental Table \ref{tab:cancers}). To assess the impact of LD on the approach, 
we applied  LD pruning of the SNPs  with  LD thresholds (e.g. pairwise LD)  0.25, 0.5 and  0.75. 
As we had access to individual level data from all studies, 
we first estimated the log-odds ratio   $\hat \beta_{sj}$ and standard error for each SNP $j$ for each cancer $s$ separately, from  logistic regression models adjusted for gender, age, study and  10 principle component scores to control for population stratification. SNPs were coded as 0, 1,or 2 minor alleles in these models. Additionally we computed phenotype-specific estimates  $\hat \Omega_s$  in  (\ref{omega}).
We then computed p-values for $T^L_{}$ and $T^Q_{}$ under the mixture model ($T^L_{Mix}$ and $T^Q_{Mix}$). For comparison, we also computed  p-values for tests $T_{Hetmeta}^L$ and  $T_{Hetmeta}^Q$  under the assumption of a single density,  given by 
\begin{equation}
\label{tlmeta}
T^L_{Hetmeta} = \frac{\sum_{s=1}^S T^L_s/Var_0(T_s^L)}{\sqrt{\sum_{s=1}^S 1/Var_0(T_s^L)}}, \end{equation} 
where $Var_0(T^L_s)$ is   calculated under the strong null model
and 
\begin{equation}
\label{tqmeta}
T^Q_{Hetmeta} = \sum_{s=1}^S \hat{\bbeta}^{*'}_{s}\Sigma^{*-2}_s\hat{\bbeta}^*_{s} =\sum_{s=1}^S T^Q_s,\end{equation} 
which is Het-MetaSKAT \citep{Tang2014,Lee2013} with weights set to 1. To test under the weak null model, we used  pancreatic cancer as a negative control.

Results from the various methods   are presented in Table \ref{tab:R1} for the LD threshold 0.5. The lowest single study p-value for the linear statistic $T^L_s$  was 0.21, observed for breast cancer. 
When we tested the strong null with the linear  $T^L_{Mix}$ and single density assumption ($T^L_{Hetmeta}$), we did not detect statistical significant associations between the genetic region and any of cancers, and the overall p-values were $1$ and $0.7$, respectively. 
In contrast, the  quadratic test $T_s^Q$ detected statistically significant association between the region  and  esophageal  cancer, with p-values 0.0001 and suggestive p-values for stomach cancer and glioma but not significant after multiple testing correction. 
Using standard meta analysis   with   $T^Q_{Hetmeta}$,  we  did not detect an overall association. {{However, $T^Q_{Mix}$ detected associations between the region and esophageal and stomach cancers, with a posterior probabilities $\hat{p}_s$ 1 and 0.61 respectively, and provided suggestive evidence for glioma ($\hat p_s =0.36$)}}. 

\begin{table}[h!]
	\centering
	\caption{{{Results of across-cancer meta analysis with fourteen cancers from case-control studies.} The number of SNPs is total number of SNPs in the region with MAF greater than 5\% and pairwise LD $<50\%$. Posterior probabilities were calculated from $T^L_{Mix}$ and $T^Q_{Mix}$ and p-values for $T^L_{Hetmeta}$ and $T^Q_{Hetmeta}$  under a single density, $\varphi_0$}.}
	\begin{tabular}{p{1.7cm}p{1.75cm}p{1.25cm}p{1.25cm}p{1.25cm}p{1.25cm}p{1.25cm}}
		&&&\multicolumn{2}{c}{Linear Test, $T^L$}&\multicolumn{2}{c}{Quadratic Test,  $T^Q$ }\\
		Cancer & Number of cases/controls & {\# SNPs} & {Posterior} & {P-value } & Posterior & {P-value }  \\
		\hline
		Bladder & {2071/6738} & {86} & 0     & 0.40 & {0.01} & 0.31  \\
		Glioma & {440/4631} & {83} & 0     & 0.50  & {0.36} & 0.08  \\
		Breast& {1035/1160} & {83} & 0     & 0.21 & {0.08} & 0.13  \\
		Colon& {109/5693} & {85} & 0     & 0.96 & {0.00} & 0.58 \\
		Endometrial & {890/713} & {79} & 0     & 0.42 & {0.01} & 0.38 \\
		Esophagus & {1956/2093} & {98} & 0     & 0.89 & {1.00} & 0.0001  \\
		Kidney & {1288/6455} & {86} & 0     & 0.98 & {0.01} & 0.87  \\
		Lung & {4786/7685} & {86} & 0     & 0.33 & {0.01} & 0.72  \\
		NHL & { 1599 /6209} & {78} & 0     & 0.63 & {0.01} & 0.65  \\
		Ovary & {278/650} & {87} & 0     & 0.31 & {0.05} & 0.60  \\
		Prostate & {5217/5043} & {82} & 0     & 0.69 & {0.01} & 0.23  \\
		Stomach & {1761/2093} & {100} & 0     & 0.34 & {0.61} & 0.02  \\
		Testis & {457/576} & {117} & 0     & 0.41 & {0.03} & 0.92  \\
		Pancreas & {417/5693} & {84} & 0     & 0.28 & {0.00} & 0.62  \\
		\hline
		&  &&  $T^L_{Mix}$   &    $T^L_{Hetmeta}$ &  $T^Q_{Mix}$ &   $T^Q_{Hetmeta}$ \\ \hline
		Global  P-value &       &       & 1     & 0.7   & 0.008 & 0.16  \\
		\hline
	\end{tabular}%
	\label{tab:R1}%
\end{table}%

 For the LD threshold 0.5, the parameters in the mixture model were $\hat{\pi}=0.2$, $\widehat{E(\mu)}=0.0018$, and  $\widehat{E(\mu_c^3)}=-0.0005$. 
 The small value of $\widehat{E(\mu)}$  indicates that signal is likely sparse in the region. We observed extremely low estimates of the heterogeneity parameters  $e^{\hat{\psi}}=1.6\times 10^{-3}$ and $e^{\hat{\zeta}}=1.5\times 10^{-8}$  because only two cancers, esophagus and stomach had a strong association with the region. The same results were observed for  SNPs selected using the LD threshold of 0.25 and 0.75 (see Supplemental Tables \ref{tab:R2} and \ref{tab:R3}). Results for stomach, esophagus cancers and glioma were previously reported to be associated with the region \citep[]{Li2014}. 

Lastly, we tested under the weak null model with $T^L_{Mix}$ and $T^Q_{Mix}$ using pancreatic cancer  as a negative control outcome. Similarly to the results for testing under the strong null model, only $T^Q_{Mix}$  under the mixture model detected the association with esophageal cancer and provided suggestive evidence for stomach cancer (Supplemental Table \ref{tab:WR1}).

\subsection{Associations of two genetic regions with expression of quantitative trait loci (eQTL) data from multiple tissues}

To illustrate our method for continuous $Y_s$, we used genotype and total gene expression data based on RNA sequencing for 17 tumor tissues from The Cancer Genome Atlas (TCGA) project. 
Details on data processing are described in Supplementary Materials of \cite{ruth2017}. Here we focused on
eQTL data from two genes,   CTSW and LARS2, and the association with  SNPs in their  cis region  (i.e.  less than 1000,000 base pairs from the target gene).

We first estimated coefficients  $\hat{\beta}_{sj}$ and  standard errors for each SNP $j$ for each tumor tissue $s$ from linear regression models,  adjusted for sex, age and the top five principle component scores, and  obtained phenotype-specific estimates $\hat{\Omega}_s$ for genotype correlations in (\ref{omega}).  We then computed  standard meta analytic tests, $T^Q_{Hetmeta}$, $T^L_{Hetmeta}$, and $T^Q_{Mix}$ and $T^L_{Mix}$ under the mixture model based on the tissue specific $\hat{\bm \beta}_s$, $\hat \Sigma_s$  and $\hat \Omega_s$.

	Results for the CTSW gene are presented in Table \ref{tab:R2a} for the LD threshold 0.5 and in Supplemental Table \ref{tab:R2Sa}  for the LD threshold 0.75. The number of cis-SNPs analyzed for the individual tissues ranged from 30 to 41. Based on $T_s^L$, the KIRC, LGG, LUSC,  UCEC tissues had  p-values $<0.05$,  but no significant associations after a multiple testing correction. When we tested using the strong null model neither  $T^L_{Hetmeta}$ nor  $T^L_{Mix}$  detected any statistical significant associations model for any of seventeen tissues. 
In contrast, $T^Q_s$ detected   statistically significant associations (even using a  Bonferroni threshold $0.05/17\approx 0.003$) with the  region for the  BLCA, BRCA, LAML, LGG, LUAD, LUSC, and  OV tissues.  
Both,    $T^Q_{Hetmeta}$ and   $T^Q_{Mix}$ detected  an overall association.
Estimated  posterior probabilities  $\hat p_s>0.75$ were observed for multiple tissues (BLCA, BRCA, KIRP, LAML, LGG, LUAD, LUSC ,OV, PRAD, and SKCM) tissues, and  suggestive evidence was provided for two tissues, UCEC and LIHC (with posterior probabilities $\hat p_s$ of  0.61 and 0.45, respectively). 
We  note that three tissues (KIRP, PRAD and SKCM) had individual study p-values $>0.003$, but  posterior probabilities $\hat p_s>0.80$ (Table \ref{tab:R2a}).  Two  of these tissues had small sample sizes,  highlighting that small studies sometimes borrow more information from the overall set of studies. We also note that the p-value from the KIRC tissue was similar to that for the PRAD tissue (both approximately equal to 0.04); however, the posterior probability estimate for this tissue was $\hat p_s\approx 0$. Our approach lessened the importance of large studies with weak evidence.
The  parameter estimates in the mixture model were $\hat{\pi}=0.61$ for the proportion of associated studies,  $\widehat{E(\mu)}=-0.0058$  and  $\widehat{E(\mu_c^3)}=6\times 10^{-5}$ for the mean genetic effect sizes,  and  $e^{\hat{\psi}}=0.008$ and $e^{\hat{\zeta}}=7\times 10^{-5}$ for the  mean values of the heterogeneity parameters. The small value of $\hat{\mu}$  indicates that the signal is   sparse and heterogeneous in the region. 
\begin{table}[t!]
	\centering
	\caption{Results from cross-tissue eQTLs association analysis with cis-SNPs, CTSW gene. The sample size is the number of measurements for the specific tissue. The number of SNPs (\#) is the  number  of SNPs in the cis region with MAF $>5\%$ and pairwise LD $<50\%$. Posterior probabilities were calculated for $T^L_{Mix}$ and $T^Q_{Mix}$ and p-values for $T^L_{Hetmeta}$ and $T^Q_{Hetmeta}$  under a single density, $\varphi_0$.}
	\begin{tabular}{p{1.7cm}p{1.25cm}p{1.25cm}p{1.25cm}p{2cm}p{1.5cm}p{2cm}}
		&&&\multicolumn{2}{c}{Linear Test, $T^L$}&\multicolumn{2}{c}{Quadratic  Test, $T^Q$}\\
		Cancers & Sample size & {\# SNPs} & {Posterior} & {P-value } & Posterior & {P-value }  \\
		\hline
		BLCA  & 266   & 37    & 0.46  & 6.08E-02 & {1} & 1.45E-05 \\
		BRCA  & 713   & 39    & 0.00  & 3.70E-01 & {1} & 1.71E-13 \\
		COAD  & 186   & 40    & 0.07  & 9.31E-01 & {0.03} & 5.79E-01 \\
		GBM   & 120   & 38    & 0.01  & 9.22E-01 & {0.06} & 8.42E-01 \\
		HNSC  & 351   & 35    & 0.00  & 2.84E-01 & {0.00} & 9.44E-02 \\
		KIRC  & 390   & 34    & 0.38  & 4.63E-02 & {0.00} & 3.98E-02 \\
		KIRP  & 92    & 32    & 0.21  & 7.13E-01 & {0.82} & 6.13E-02 \\
		LAML  & 154   & 30    & 0.42  & 2.12E-01 & {1.00} & 6.78E-04 \\
		LGG   & 326   & 36    & 0.28  & 4.53E-02 & {1.00} & 2.88E-04 \\
		LIHC  & 75    & 41    & 0.20  & 9.43E-01 & {0.45} & 2.69E-01 \\
		LUAD  & 427   & 33    & 0.01  & 9.03E-02 & {1} & 7.28E-08 \\
		LUSC  & 407   & 38    & 0.02  & 5.83E-03 & {1} & 1.15E-08 \\
		OV    & 219   & 36    & 0.22  & 8.02E-02 & {1.00} & 2.94E-03 \\
		PAAD  & 149   & 36    & 0.15  & 8.70E-01 & {0.03} & 7.74E-01 \\
		PRAD  & 153   & 39    & 0.30  & 7.21E-01 & {0.84} & 3.76E-02 \\
		SKCM  & 354   & 40    & 0.11  & 7.45E-02 & {0.97} & 5.72E-03 \\
		UCEC  & 268   & 39    & 0.61  & 4.85E-02 & {0.61} & 1.15E-01 \\
		\hline
		&  &&  $T^L_{Mix}$   &    $T^L_{Hetmeta}$ &  $T^Q_{Mix}$ &   $T^Q_{Hetmeta}$ \\ \hline
		Global  P-value&       &      & 0.50  & 3.37E-01 & $<$0.001 & 2.81E-14 \\
		\hline
		\multicolumn{7}{p{12.5cm}}{
			BLCA:  Bladder Urothelial Carcinoma; BRCA: Breast invasive carcinoma; COAD: Colon adenocarcinoma; 
			GBM: Glioblastoma multiforme; HNSC: Head and Neck squamous cell carcinoma;KIRC: Kidney renal clear cell carcinoma; KIRP: Kidney renal papillary cell carcinoma; LAML: Acute Myeloid Leukemia; 
			LGG: Brain Lower Grade Glioma; LIHC: Liver hepatocellular carcinoma;
			LUAD: Lung adenocarcinoma; LUSC: Lung squamous cell carcinoma;
			OV: Ovarian serous cystadenocarcinoma; PAAD: Pancreatic adenocarcinoma;
			PRAD: Prostate adenocarcinoma; SKCM: Skin Cutaneous Melanoma; 
			UCEC: Uterine Corpus Endometrial Carcinoma} 
	\end{tabular}%
	\label{tab:R2a}%
\end{table}

Results for the eQTL data for the seventeen tissues and SNPs from the LARS2 gene are presented in Supplemental Tables \ref{tab:R2b} and \ref{tab:R2Sb}. The linear tests $T^L_{Hetmeta}$ an $T^L_{Mix}$ did not detect an association between tissues and LARS2, while both quadratic testd did. Based on $T^Q$, the posterior probabilities for all tissues were equal to one. 
The  parameter estimates in the mixture model were $\hat{\pi}=1$ for the proportion of associated studies,  $\widehat{E(\mu)}=0.03$  and  $\widehat{E(\mu_c^3)}=0.00002$ for the mean values of the genetic effect sizes,  and  $e^{\hat{\psi}}=0.001$ and $e^{\hat{\zeta}}=8.5\times 10^{-4}$ for the  mean values of the heterogeneity parameters. Large values of these parameters indicate that a single density with heavy tails is the best fit to the data. Therefore, our approach may have lower specificity when the proportion of associated studies and estimated effects are heterogeneous as indicated by a large posterior probability for the PAAD tissue, which  had a marginal p-value of 0.41. 

For this example, we did not test under the weak null model as we did not have knowledge about a negative control study. 

\section{Simulations}

\subsection{Setup}

We assessed the type 1 error and the power of the mixture method for both binary and continuous outcomes, $Y_s$. To generate realistic patterns of LD,  we used genotypes of common SNPs (MAF$\geq$ 5\%) on chromosome 6 observed in the 4631 controls from the glioma  study \citep{brain} also used in Section 4.1.  We applied LD pruning to ensure that the maximal pairwise LD between SNPs was no larger than 0.5.
For each setting  we generated $S=20$ studies, of which $S_C=0,1,5,10$ and $15$ studies had  SNPs associated with $Y_s$.
We investigated   two LD patterns.  For the  ``high LD pattern'' setting  we used genotypes for 210 common SNPs in the region from 29600054bp to 31399945bp on chromosome 6 (HLA I class region).  For the ``low LD pattern'', we selected $p=210$ SNPs in the region from 110391bp to 1525603b on chromosome 6  with pairwise LD smaller than 0.5.
We also studied the impact of sample size of the studies with no signal on power. For binary $Y_s$, the sample size for studies with truly associated SNPs was  $N^1_{case}=N^1_{cont}=2500$, and the sample sizes of studies with no signal was  $N^0_{case}=N^0_{cont}=3500$,  $N^0_{case}=N^0_{cont}=2500$ and  $N^0_{case}=N^0_{cont}=1500$.  For continuous outcomes, the sample size for studies with causal SNPs was  $N^1=5000$, and for studies with no signal was  $N^0=7000$,  $N^0=5000$ and  $N^0=3000$.

For studies under the strong null model, we generated $N=5000$ phenotypes $Y_{si}$ from $N(0,1)$  and for binary $Y_s$, we randomly assigned  2500 cases and 2500 controls to 5000 genotypes. For the $S_c$ studies with truly associated SNPs, we randomly selected $p_C=11$ of the $p=210$ SNPs and  generated $\gamma_{sj}$ for $j=j_1,...,j_{11}$ in model (\ref{model})  from generated $N(\mu_{sj},\tau_{sj})$, where $\mu_{sj} \sim N\left\{E(\mu),(E(\mu)/4)^2\right\}$ and  $\tau_{sj} \sim TN\left\{E(\tau),(E(\tau)/2)^2\right\}$ where  $TN$ denotes a normal distribution truncated at 0.
Continuous phenotypes were generated from $Y_{is} = \bm \gamma_s' \bm X_{is} +e_{si}$, where $e_{si} \sim N(0,1)$. For simulations based on case control data, we   generated  $Y_s \sim Bernoulli(q_s)$, where $q_s = \exp({\gamma_0+\bm \gamma'_s\bm X_{is}})/\left\{1+\exp({\gamma_0+\bm \gamma'_s\bm X_{is}})\right\}$ with $\gamma_0=log(0.01/0.99)$ for a large cohort and then sampled cases and controls.

Under the weak null model for  null SNPs,  we generated $\gamma_{sj}$ from $\gamma_{sj} \sim N(0,\tau_{sj})$ and $\tau_{sj} \sim TN\left\{E(\tau),(E(\tau)/2)^2\right\}$ for $s=1,...,20$. For the $S_C$ studies with  $p_C=11$  randomly selected truly associated SNPs,
we generated $\gamma_j$ for $j=j_1,...,j_{11}$ using the hierarchical structure in Figure \ref{Figure1}.

For both, the strong and weak null models, we investigated the type 1 error ($S_C=0$) and the power of $T^L_{Mix}$ and $T^Q_{Mix}$ for $S_C=1,5,10$ and $15$. We used  two estimates for the matrix $\Omega_s$  of correlations among SNPs,   in (\ref{omega}): 1) a global external estimate obtained from the original 4250 original controls and 2) and internal estimates obtained separately  for each study ($\hat{\Omega}_s$, $s=1,...,20$) based on the observed genotypes.

We compared the power of $T^L_{Mix}$ and $T^Q_{Mix}$ to  that of $T^L_{Hetmeta}$ and $T^Q_{Hetmeta}$   in $(\ref{tlmeta})$ 
and   $(\ref{tqmeta})$, and additionally to   the sum of Hotelling  tests,  
$T_{Hotmeta} = \sum_{s=1}^S \hat{\bbeta}^{*'}_{s}\Sigma^{*-1}_s\hat{\bbeta}^*_{s}.$ The asymptotic distributions for these tests are calculated under the strong null model \citep[]{Tang2014,Lee2013}.  
For $T^L_{Hetmeta}$ and $T^Q_{Hetmeta}$, we used a LRT similar to that used for the mixture models for $T^L_{Mix}$ and $T^Q_{Mix}$, but with $\pi=1$ under the  alternative model.

\subsection{Simulation results}
\subsubsection{Type 1 error for testing for heterogeneity of associations}
The empirical type 1 error rates for our $T^L_{Mix}$ and $T^Q_{Mix}$   with binary and continuous outcomes are presented in detail in Supplemental Tables \ref{type1} and \ref{type1_cont} for $\alpha=0.05$ and $\alpha=0.01$, for testing under the strong and weak null models. The mixture model with $T^L_{Mix}$ had the nominal  type 1 error,  regardless of the LD pattern, type of estimate of  $\Omega_s$ or type of null model. For the mixture model with $T^Q_{Mix}$ when LD was low, the empirical type 1 error was slightly conservative for both internal and external estimates of $\Omega_s$. However, when LD was high, the empirical type 1 error estimates were more conservative for both null models for external estimates of $\Omega_s$ that do not capture LD patterns as accurately as internally estimated $\Omega_s$. {Overall our empirical results confirm that the type 1 error is controlled when $p_s$ is large.}

\subsubsection{Power to test under the strong null model}

\begin{figure}[t!]
	\centering
	\begin{tabular}{p{0.005cm}ccc}
		A:  & \adjustbox{valign=m,vspace=0.5pt}{ \includegraphics[width=.3\linewidth]{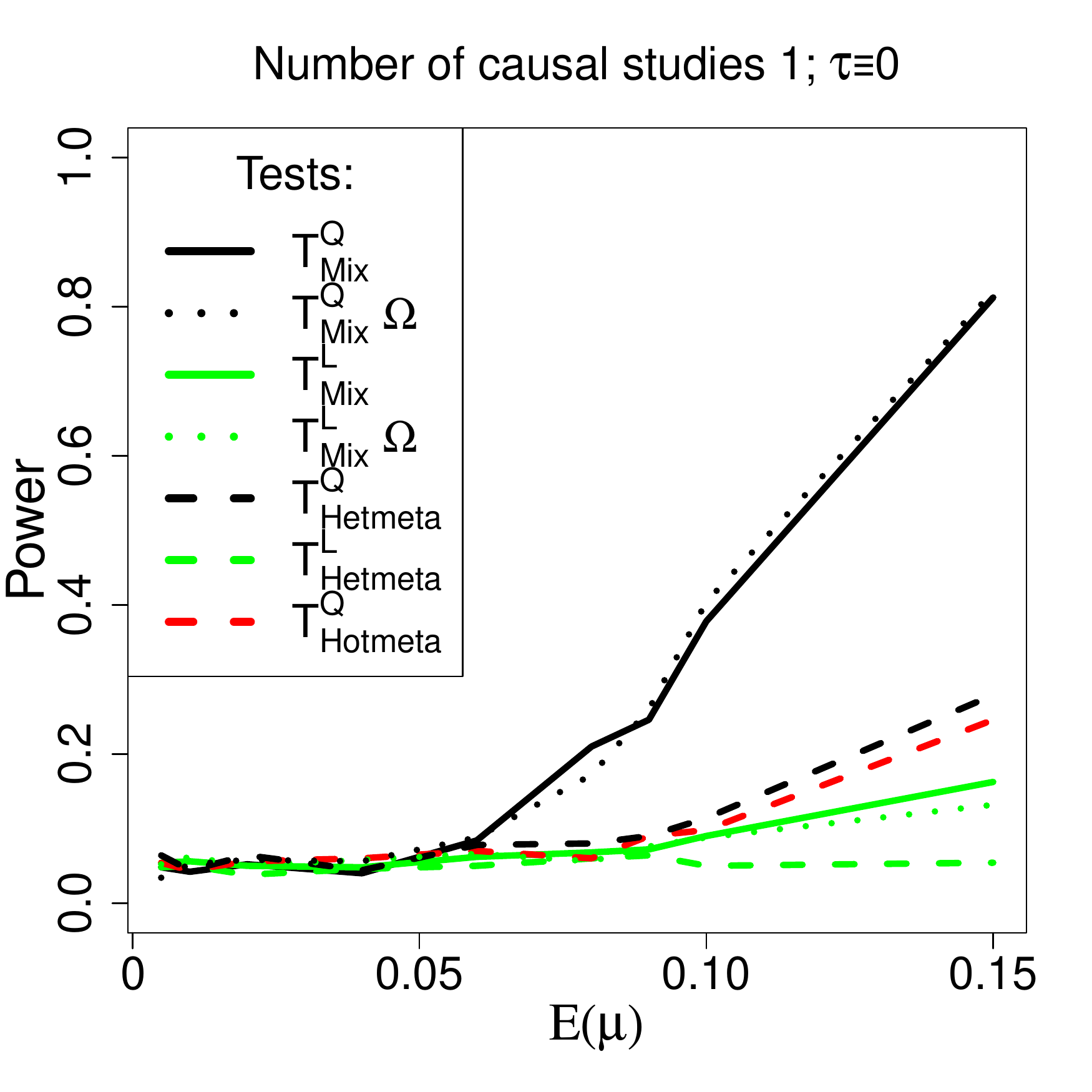}} & \adjustbox{valign=m,vspace=0.5pt}{ \includegraphics[width=.3\linewidth]{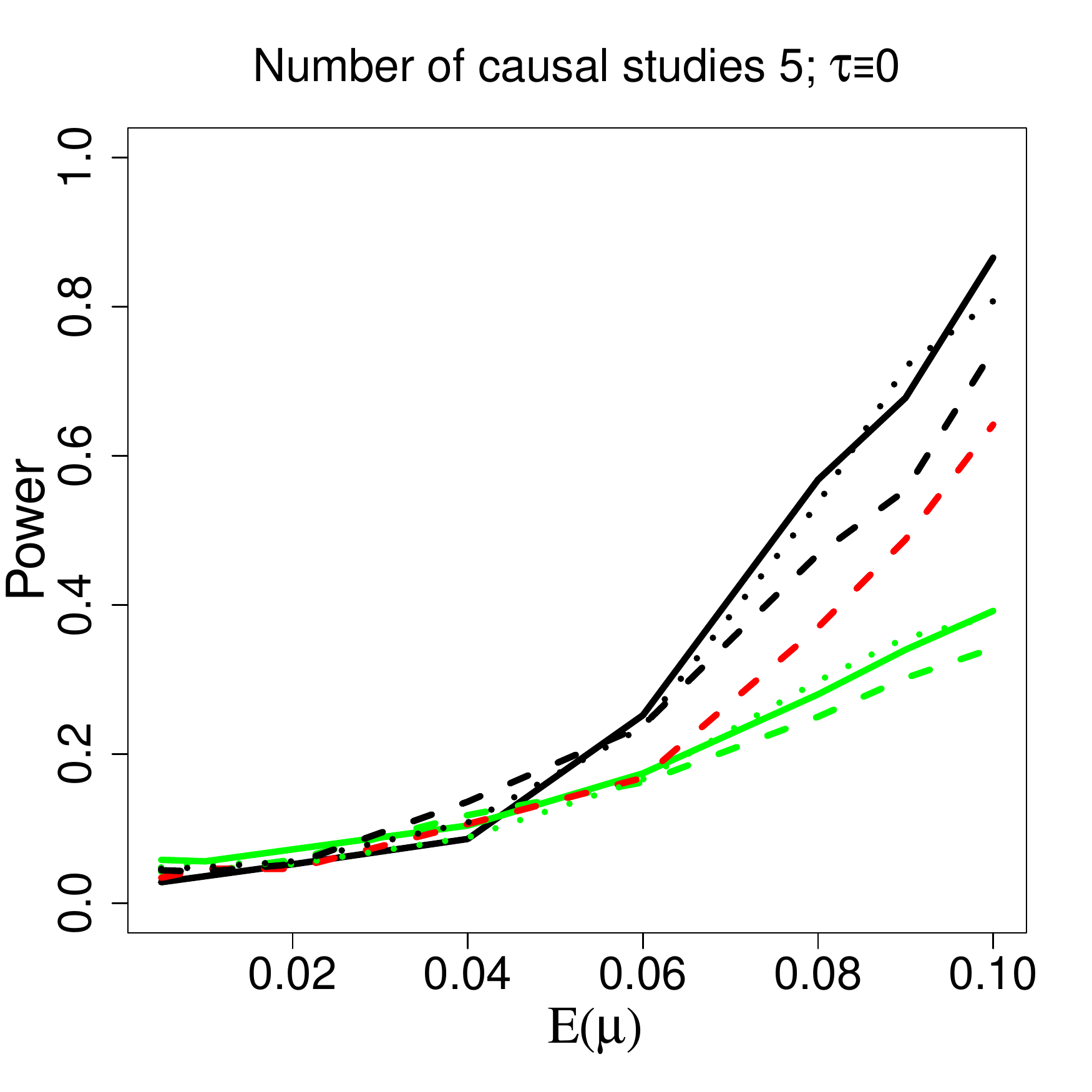}}  & \adjustbox{valign=m,vspace=0.5pt}{ \includegraphics[width=.3\linewidth]{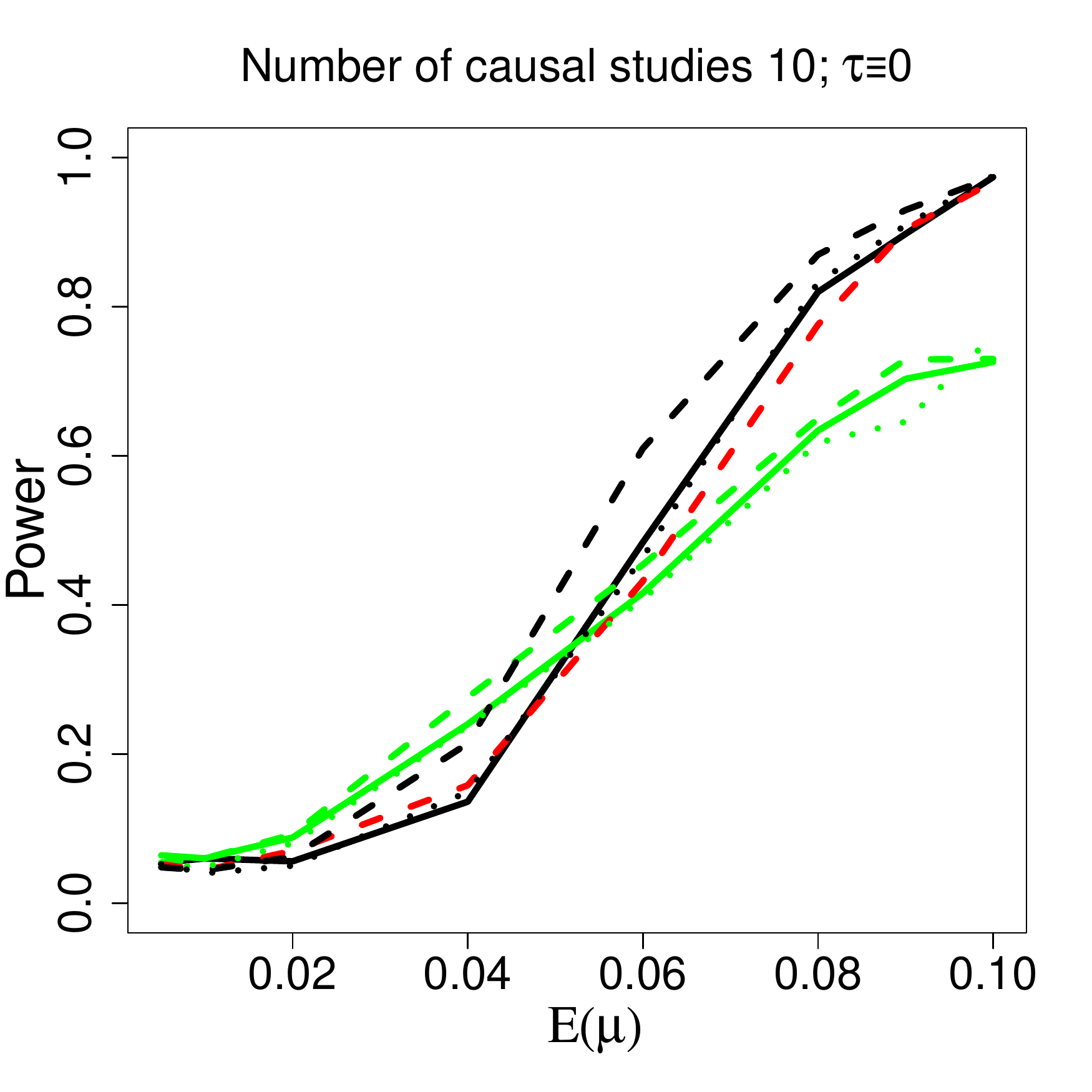}}  \\
		B:  & \adjustbox{valign=m,vspace=0.5pt}{ \includegraphics[width=.3\linewidth]{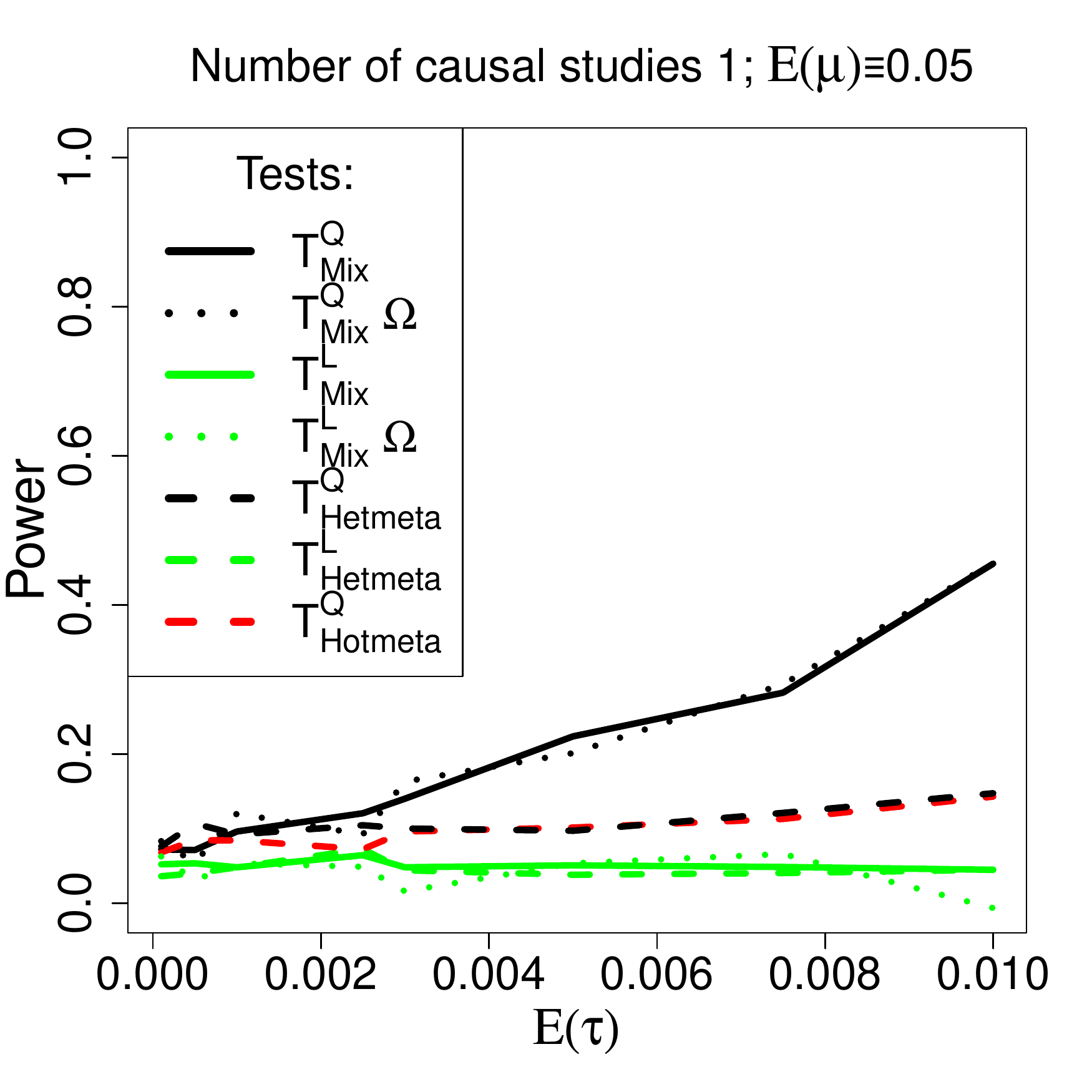}} & \adjustbox{valign=m,vspace=0.5pt,}{ \includegraphics[width=.3\linewidth]{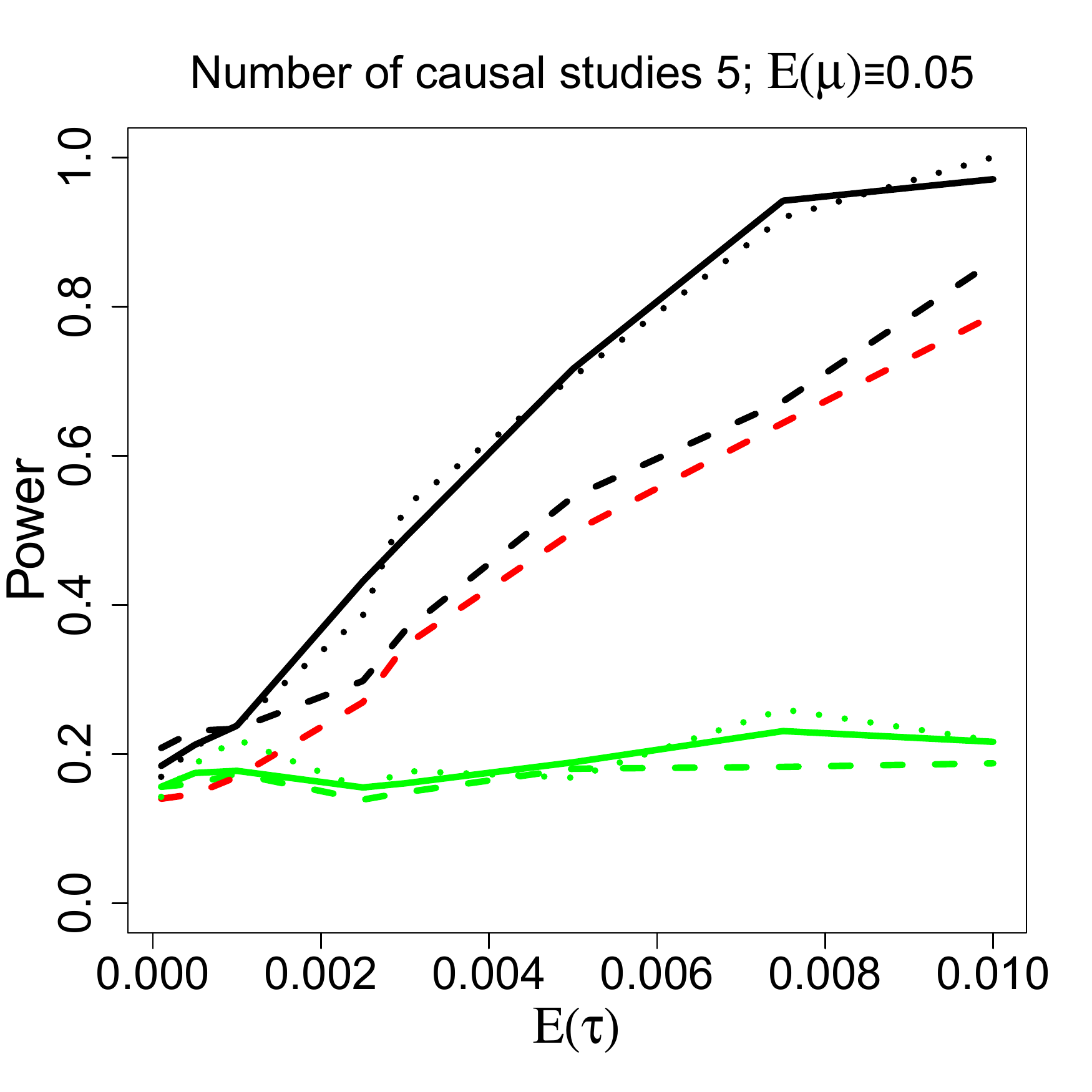}}  & \adjustbox{valign=m,vspace=0.5pt}{ \includegraphics[width=.3\linewidth]{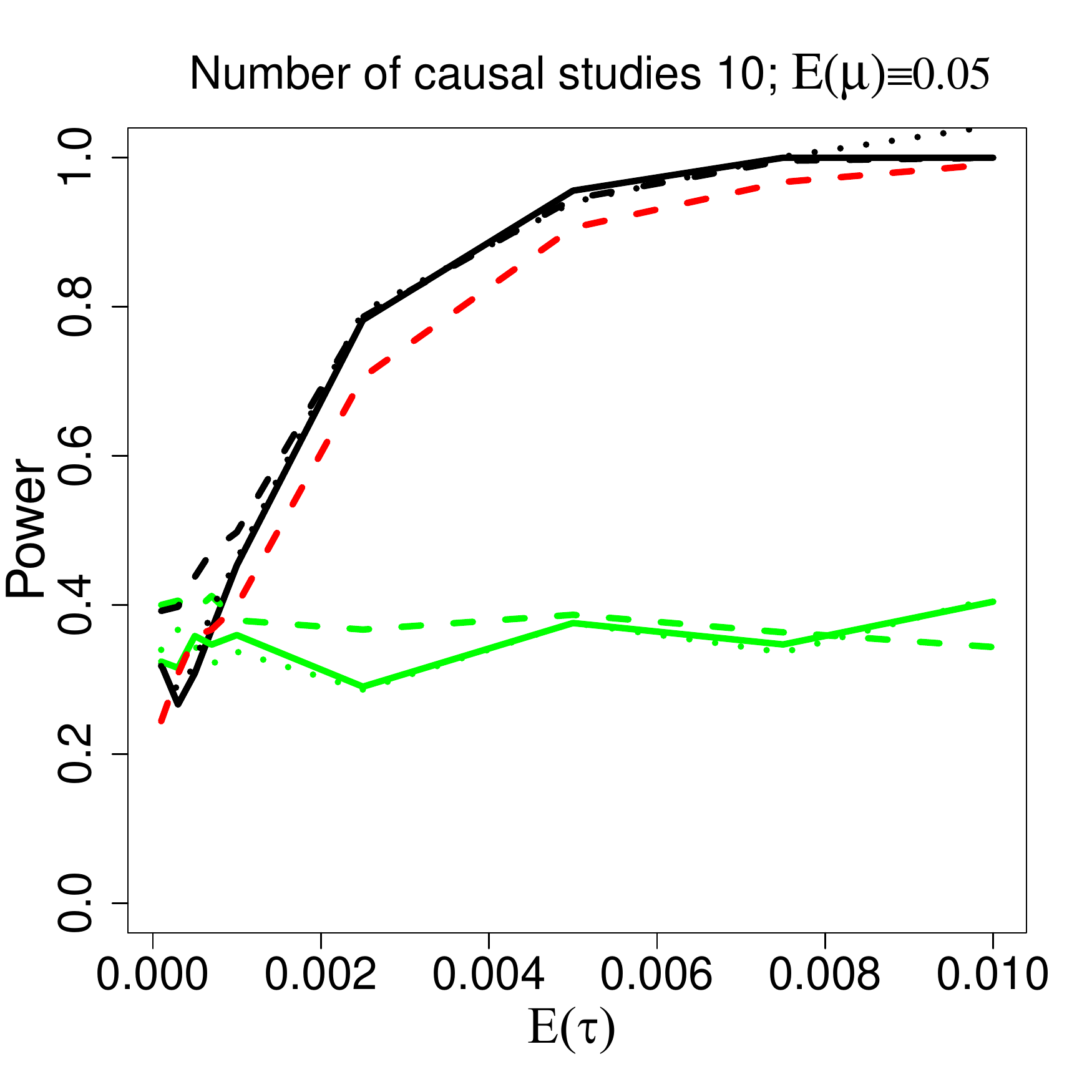}}  
	\end{tabular}
		\caption[]{ Empirical power comparisons for binary phenotypes between various methods for testing the strong null model ($\bm \mu\equiv \0, \bm \tau \equiv \0$).  Low LD pattern between SNPs in a region; sample sizes per study without and with signal are $N^0_{case} = N^0_{cont} = 3500,~N^1_{case}=N^1_{cont} = 2500$. $T^Q_{Mix}~\Omega$ and $T^L_{Mix}~\Omega$ use external estimate of matrix $\Omega$. Level of the test is 0.05 and $S=20$ studies are analyzed. \textbf{Panel A}: SNP effects under alternative $E(\mu)\neq 0$ and $\bm \tau \equiv \0$;\textbf{ Panel B:} SNP effects under alternative $E(\mu)=0.05$ and $E(\tau) \neq 0$.}
		        \label{Fig1}
		        \label{Fig2}
\end{figure}

Here, we focus on findings for binary $Y_s$. Results for continuous $Y_s$ were qualitatively similar and are presented  in Supplemental Figures \ref{FigC1} and \ref{FigC4}. The results from our power studies are summarized in Figure \ref{Fig1}, Supplemental Figures \ref{Fig4} - \ref{Fig14}. The mixture approach had better power than  other methods (Figure \ref{Fig1},  Supplemental Figures \ref{Fig4} - \ref{Fig14} ) when the proportion of studies with associated SNPs was below 50\%. When the  proportion of studies with signal was above 50\%, $T^L_{Hetmeta}$ and $T^Q_{Hetmeta}$ had better power than $T^L_{Mix}$ and $T^Q_{Mix}$ (Supplemental Figures \ref{Fig4} - \ref{Fig14}). For the same settings, the linear tests, $T^L_{Mix}$ and $T^L_{Hetmeta}$ had higher power when effect sizes were small and in the same directions (Supplemental Figures \ref{Fig3} - \ref{Fig10}). But, as expected they were not powerful when the genetic effects were heterogeneous (Figure \ref{Fig2} and Supplemental Figures \ref{Fig3} - \ref{Fig14}). The empirical power of  $T^Q_{Mix}$ and $T^L_{Mix}$ was not noticeably affected by the sample size of studies not associated with the region. Similarly, $T_{Hotmeta}$ was not affected  by the sample size of the null  studies, because it  explicitly assigns the same weight to each study. In contrast, the power of $T^Q_{Hetmeta}$ was higher when the sample sizes of the studies with associated SNPs were larger than those with no signal (Supplemental Figures  \ref{Fig7} - \ref{Fig10}). We saw similar results for the eQTL data and SNPs from the LARS2 gene (Section 4.2)
When LD in a region was high, using external estimates for $\Omega_s$ resulted in more conservative Type 1 error and thus  decreased of power of $T^Q_{Mix}$ and slightly lower power of $T^L_{Mix}$. When $\Omega_s$ was estimated from study specific data, the power of tests was similar regardless of LD  pattern (Figure \ref{Fig1}, Supplemental Figures \ref{Fig4} - \ref{Fig14} ). 

\subsubsection{Power to test under the weak null model}

\begin{figure}[t!]
	\centering
	\begin{tabular}{p{0.01cm}ccc}
		A:  & \adjustbox{valign=m,vspace=0.15pt}{ \includegraphics[width=.28\linewidth]{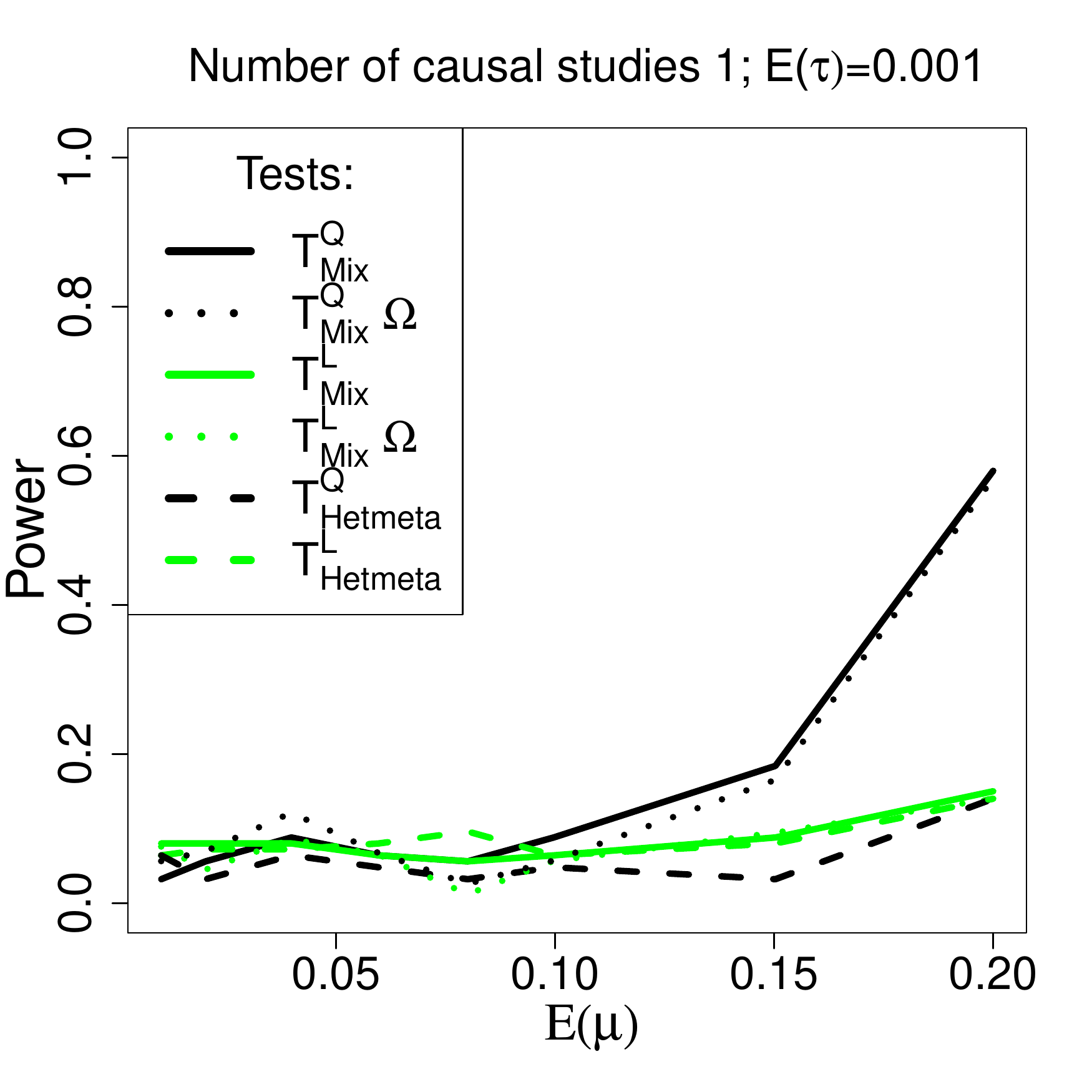}} & \adjustbox{valign=m,vspace=0.15pt}{ \includegraphics[width=.28\linewidth]{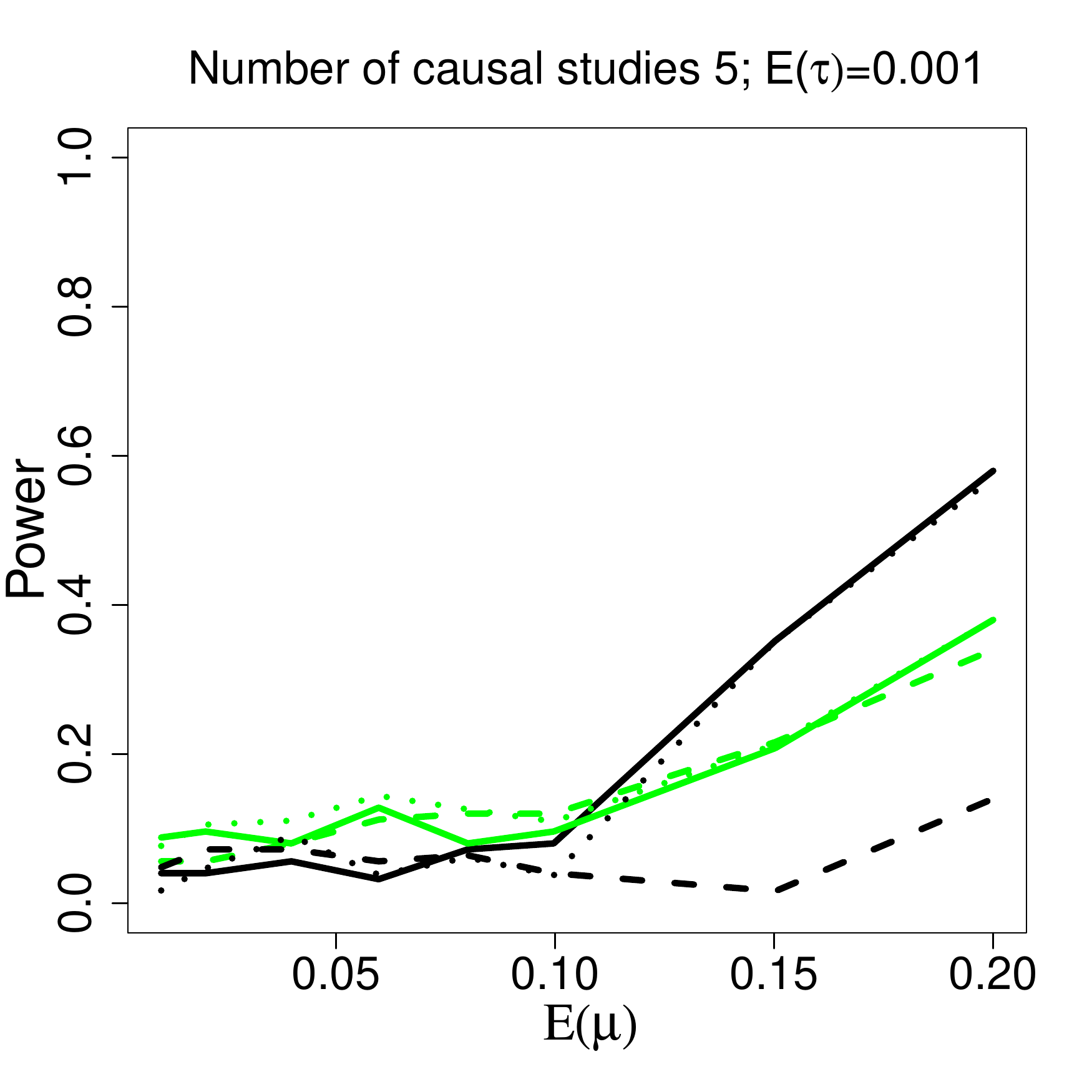}}  & \adjustbox{valign=m,vspace=0.15pt}{ \includegraphics[width=.28\linewidth]{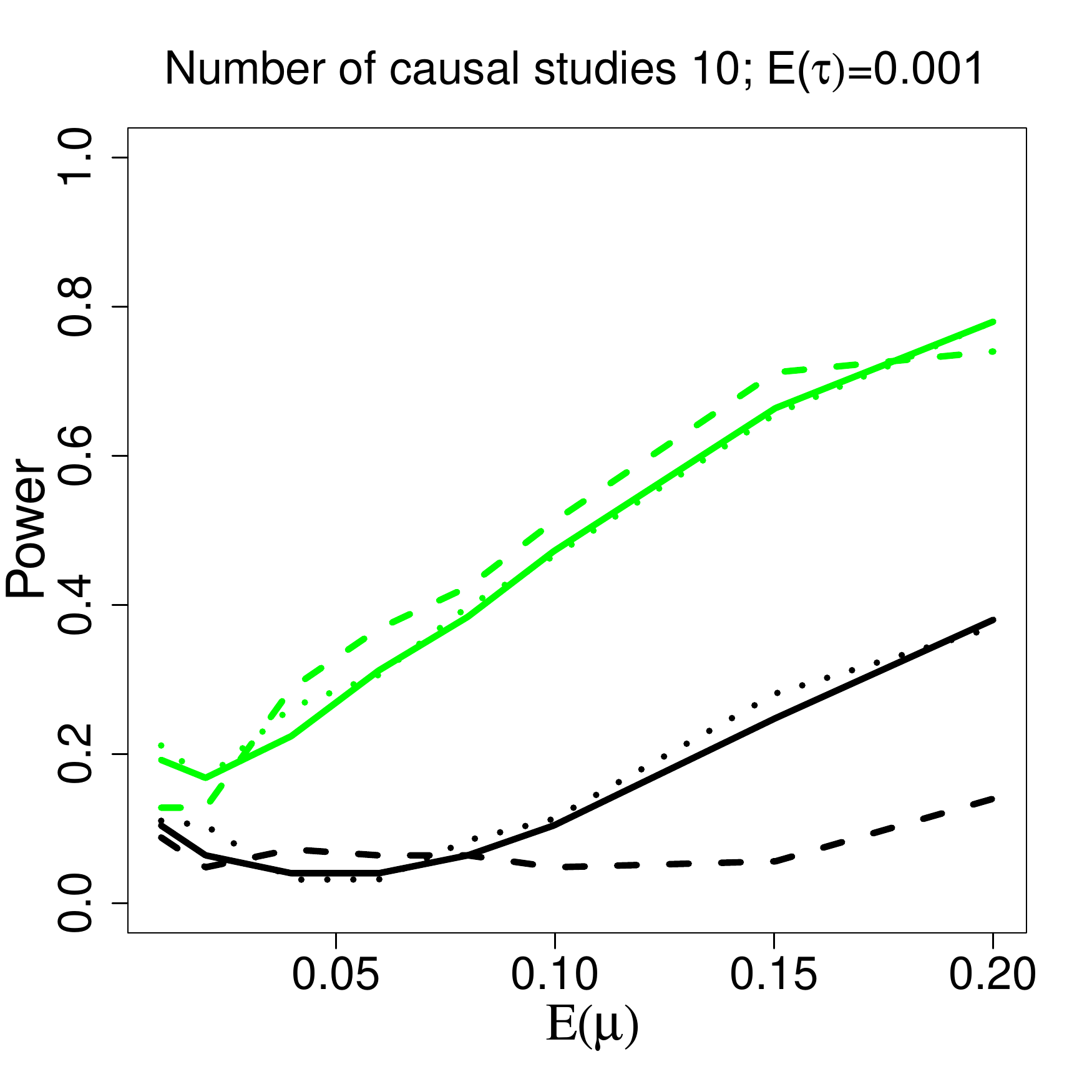}}  \\
		B:  & \adjustbox{valign=m,vspace=0.15pt}{ \includegraphics[width=.28\linewidth]{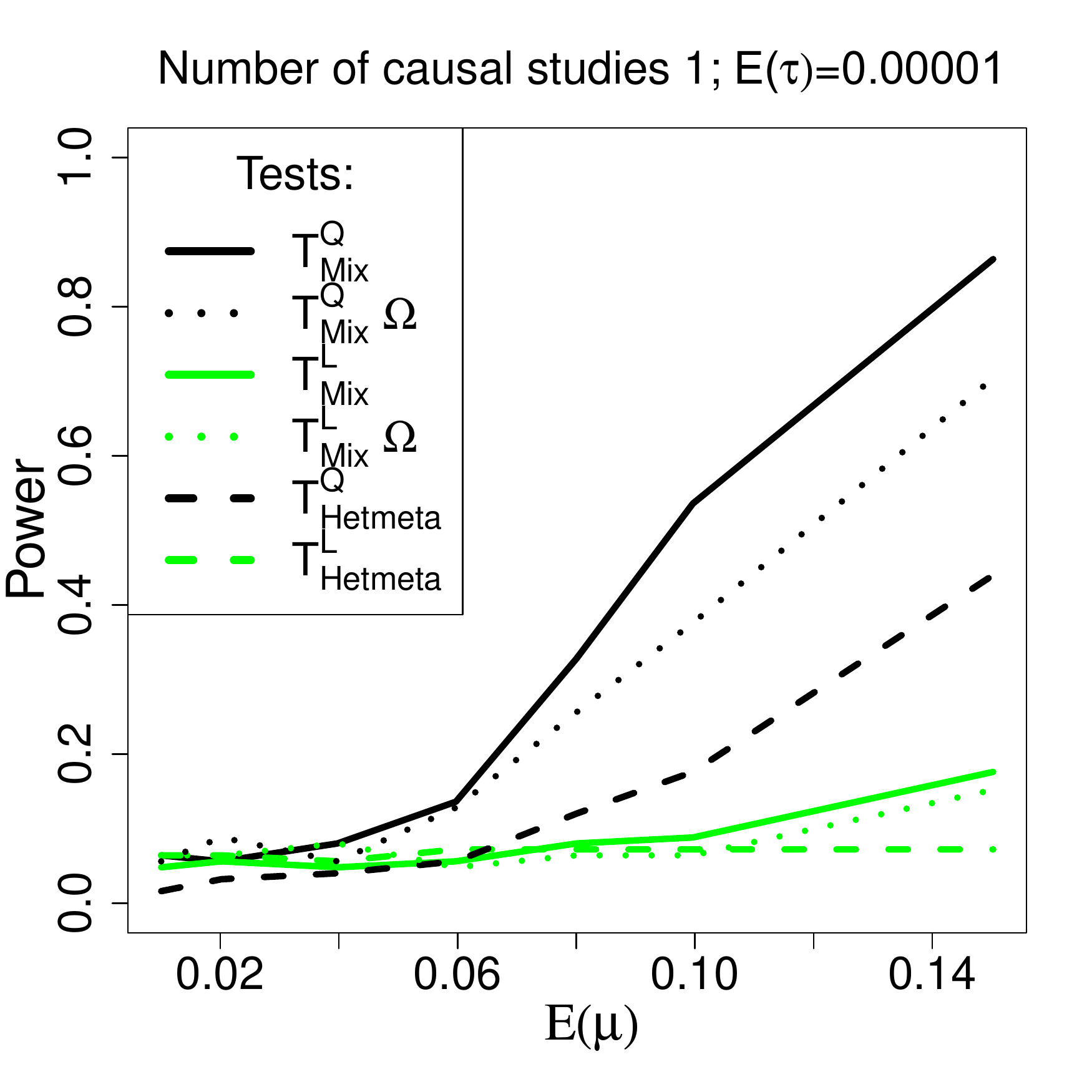}} & \adjustbox{valign=m,vspace=0.15pt}{ \includegraphics[width=.28\linewidth]{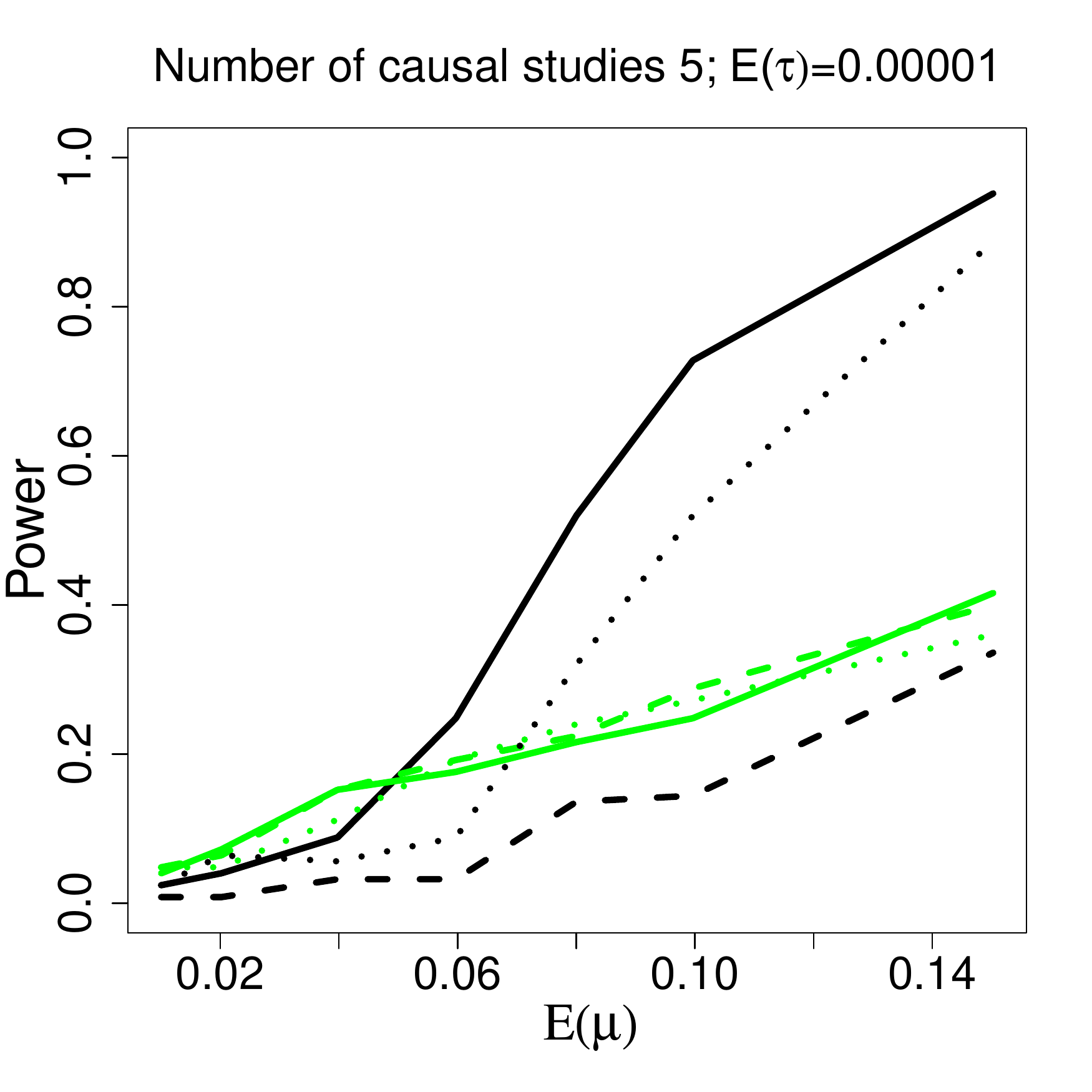}}  & \adjustbox{valign=m,vspace=0.15pt}{ \includegraphics[width=.28\linewidth]{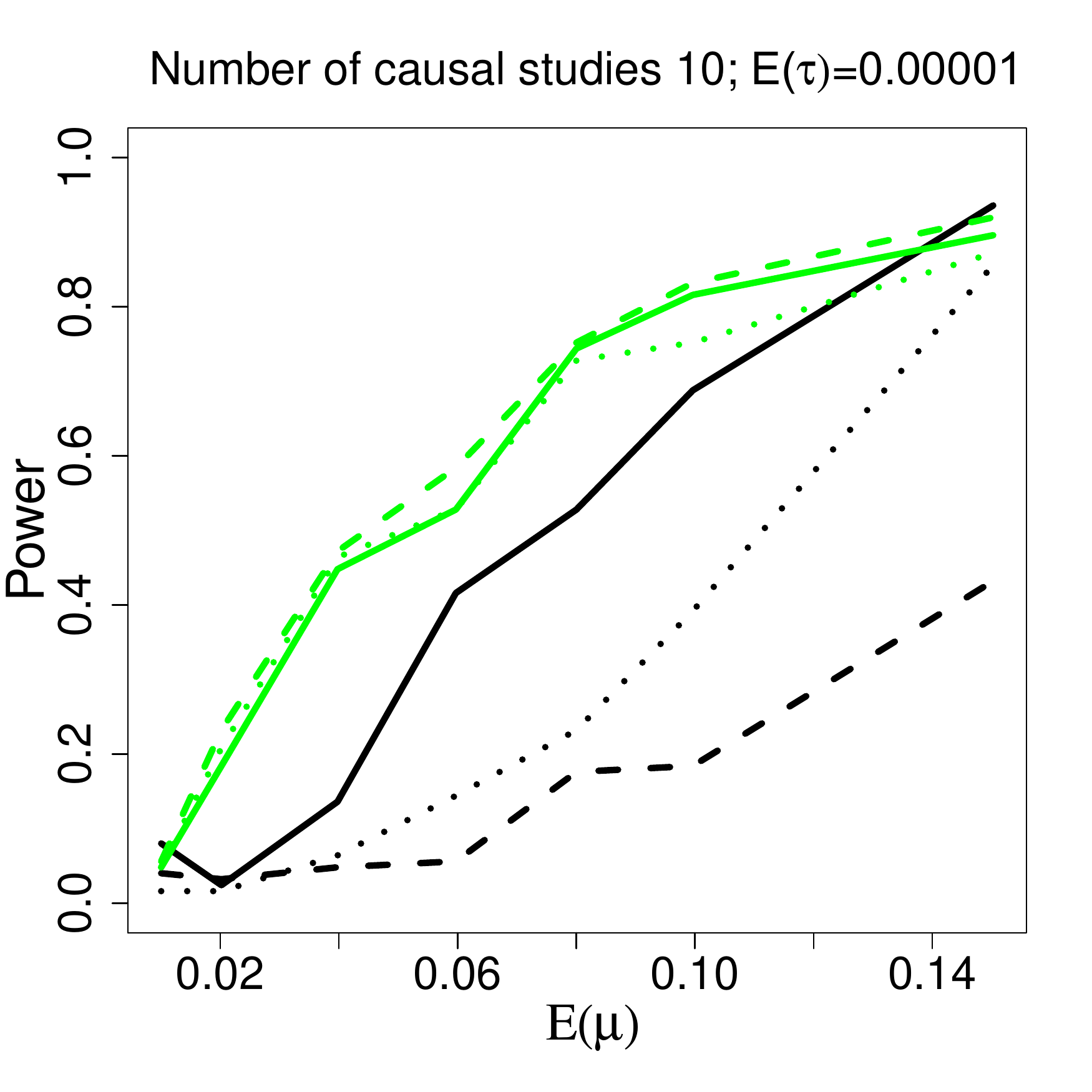}}  
	\end{tabular}
	\caption[]{Empirical power comparisons for binary phenotypes between various methods for testing the weak null model ($\bm \mu\equiv \0$). Sample sizes per study without and with signal are $N^0_{case} = N^0_{cont} = 3500,~N^1_{case}=N^1_{cont} = 2500$. $T^Q_{Mix}~\Omega$ and $T^L_{Mix}~\Omega$ use external estimate of matrix $\Omega$. Level of the test is 0.05 and $S=20$ studies are analyzed. \textbf{Panel A}: Low LD pattern between SNPs in a region\textbf{ Panel B:} High LD pattern between SNPs in a region.}	
	\label{Fig5}
	\label{Fig6}
\end{figure}

Similarly to  testing under the strong null model, the LD pattern did not noticeably impact the power when $\Omega_s$ was estimated internally (Figure \ref{Fig5}, Supplemental Figures \ref{Fig15} and \ref{Fig18}). The quadratic test statistic $T^Q_{Mix}$ had higher power than all other tests when the proportion of studies with associated  SNPs was small (Figure \ref{Fig5}). However, the power of $T^Q_{Mix}$ dropped noticeably when more than 50\% studies had associated SNPs, was depended on sample sizes of the studies with causal SNPs (Figure \ref{Fig5}, Supplemental Figures \ref{Fig15} and \ref{Fig18}). The reason for this loss of power is that when $E(\mu)$ is large,   the variance of the component density $\varphi_1$ of the mixture (\ref{mixture}) is  much larger than the variance for  $\varphi_0$, which makes it challenging to identify heterogeneity of associations, as a single component density may fit the observed data as well as the mixture.  Similarly,  $T^Q_{Hetmeta}$ had low power for  all simulation scenarios under the  weak null (Figure \ref{Fig5}). The power of the linear tests under both, the mixture and single component density models,  increased as the number of studies with signal increased (Figure \ref{Fig5}), because
$\varphi_0$ has mean equal to 0 under the null model. Lastly,  the power for testing under the weak null model was much higher when the null studies had larger sample sizes, because they provide more  information on the true amount of heterogeneity captured by $\tau$.


\section{Discussion}

We proposed a novel approach based on a mixture model to  assess the heterogeneity of associations of genetic variation in a pre-specified region with different phenotypes, and to identify the subset of  phenotypes  associated  with the region. 
Our simulations and a data example using eQTL data show that when the proportion of associated phenotypes is less than $50\%$, combining region specific estimates using a quadratic test statistic under the mixture model assumption had much better power to identify truly associated outcomes than standard meta analytic approaches. However, when the proportion of associated outcomes was high, standard meta analytic methods were more powerful than our approach. Similar conclusions were previously reached  in the context of testing rare variants, where  
using  linear tests with  data driven weights 
worked well when the  proportion of variants with signal  was low, but a simple sum test had better power when the proportion was high  \citep{derkach2014}.

There are many  tests for associations between a genetic region and  a single phenotype  for common  \citep[e.g.][]{Zaykin2002,Sluis2015}
and rare SNPs \citep[e.g.][]{  	
	Neale2011, Lee2012}. Aggregated level methods for common variants for testing gene- and pathway level associations typically are based on p-values \citep{Sluis2015}. 
Few methods exist to assess cross-phenotype associations using summary statistics.
\citet{Bhattacharjee2012}  extended fixed effects meta analysis for a single SNP by allowing some subsets of outcomes to have no associations. Our method expands this work in two ways. First, we aggregate association estimates from multiple SNPs measured in a region, and thus utilize information stemming from LD. We also quantify heterogeneity between associations for different phenotypes. Another advantage of our  approach is that  it   allows one to incorporate prior or  external information on the likelihood that a phenotype exhibits associations with a region via the mixing proportion, which can improve identification of associated outcomes.  
Our  framework also extends a recently proposed Bayesian method (CPBayes) for testing the association between a single SNP and multiple phenotypes \citep{Majumdar2017}. CPBayes imposes a spike and slab prior on the genetic SNP effect and uses a mixture of two normal distributions to represent the SNP effect  under the  null and alternative models.  
When a single SNP is analyzed, our mixture set up corresponds to that of CPBayes.  However, we additionally estimate the amount of heterogeneity between outcome specific associations, captured by the parameter $\tau$,  directly from the data, while  in \citet{Majumdar2017}   it is pre-specified. Mis-specifying the amount of heterogeneity will lower power,  sensitivity and specificity of the procedure in \citet{Majumdar2017}.

Our approach  also differs from other recently proposed methods for gene-based testing  that require  phenotypes to be measured on the same individuals to  estimate between phenotype correlations \citep{Sluis2015,Tang2012,Kwak2017}. For cancer outcomes  one could simply assume outcomes are uncorrelated, as it is exceedingly unlikely to be diagnosed with two primary cancers and apply these methods to the summary statistics from multiple studies to test whether there is at least one study that shows associations. However,  these methods cannot  identify  which particular outcomes are associated with the SNPs in a gene/region. 

Our work extends beyond testing the presence of any association between SNPs in a region for multiple outcomes. Using  the weak null model, we also assess if associations are due to common signal or due to  heterogeneity. A limitation is that to test under the weak null model, we require availability  of a study without association. This control phenotype study helps  distinguish between-study heterogeneity   from true underlying associations.
Another limitation of our method is that if study specific estimates of $\Omega_s$ are   not  available,   one needs to use publicly available genetic data  such as  \citet{1000G} to estimate $\Omega_s$, which results in somewhat including lower power.

Several problems remain to be addressed in future work, handling shared controls between studies and more efficient permutation approaches to compute p-values for our model. 

\section{Software}
	\label{sec5}
	
	Software in the form of R code, together with a sample
	input data set and complete documentation is available at \\ https://github.com/derkand/STAMP.
\section*{Acknowledgments}
We used the computational resources of the NIH HPC Biowulf cluster and  thank  Drs. Mitch Gail and   Josh Sampson  for  helpful comments.

{
	\bibliographystyle{biorefs}
	\bibliography{references2016}
}

\end{document}